\documentclass[12pt]{article}

\usepackage{amssymb,amsfonts,amsmath}
\usepackage{longtable}
\usepackage{graphicx}
\usepackage{epsfig}
\usepackage{subfig}
\usepackage{url}

\newtheorem{theorem}{Theorem}[section]

\newtheorem{lemma}{Lemma}[section]

\begin{document}

\title{Deciding whether follow-up studies have replicated findings in a preliminary large-scale ``omics' study"}
\maketitle

\begin{center}
Ruth Heller \\
\emph{Department of Statistics and Operations Research, Tel-Aviv
university, Tel-Aviv, Israel. E-mail: ruheller@post.tau.ac.il}\\
Marina Bogomolov \\
\emph{Faculty of Industrial  Engineering and Management, Technion --
Israel Institute of Technology, Haifa, Israel. E-mail:
marinabo@tx.technion.ac.il }\\
Yoav Benjamini \\
\emph{Department of Statistics and Operations Research, Tel-Aviv
university, Tel-Aviv, Israel. E-mail: ybenja@post.tau.ac.il}\\
\end{center}

\begin{abstract}
We propose a formal method to declare that findings from a primary study have been replicated in a follow-up study.
Our proposal is appropriate for primary studies that involve large-scale searches for rare true positives (i.e. needles in a haystack).
Our proposal assigns an $r$-value to each finding; this is the lowest false discovery rate at which the finding can be called replicated.
Examples are given and software is available.
\end{abstract}

{\em The use of big data is becoming a central way of discovering knowledge in modern science. Large amounts of potential findings are screened in order to discover the few real ones. In order to verify these discoveries a follow-up study is often conducted, wherein only the promising discoveries are followed-up. This is a common research strategy in genomics, proteomics and in other areas where high throughput methods are used.
We show how to decide whether promising findings from the preliminary study are replicated by the follow-up study, keeping in mind that the preliminary study involved an extensive search for rare true signal in a vast amount of noise. The proposal computes a number, the $r$-value, to quantify the strength of replication.
}
\vspace{0.5cm}

We are concerned with situations in which many features
are scanned for their statistical significance in a primary study.
These features can be single nucleotide polymorphisms (SNPs) examined for associations with disease,
genes examined for differential expression, pathways examined for enrichment, protein pairs examined for protein-protein
interactions, etc. Interesting features are selected for
follow-up, and only the selected ones are tested in a
follow-up study.

This approach addresses two goals. The first goal is to increase the number of cases in order to increase the power to detect a feature, at a lower cost.
The second goal is to address the basic dogma of science that a finding
is more convincingly a true finding if it is replicated in at least one more study.
Replicability has been the cornerstone of science as we know it since the foundation of experimental science.  Possibly the first documented example is the discovery of a phenomenon related to vacuum, made by Huygens in Amsterdam in the 17th century, who travelled to Boyle's laboratory in London in order to replicate the experiment and prove that the scientific phenomenon was not idiosynchronic to his specific laboratory with his specific equipment \cite{Shapin85}. In modern research, the lack of replicability has deeply bothered behavioral scientists that compare the behavior of different strains of mice, e.g. in knockout experiments.  It is well documented that in different laboratories, the comparison of behaviors of the same two strains may lead to opposite conclusions that are both statistically significant (\cite{Crabbe99}, \cite{kafkafi05}, and Chapter 4 in \cite{Crusio13}).  An explanation may be the different laboratory environment (i.e. personnel, equipment, measurement techniques)  affecting differently the study strains (i.e. an interaction of strain with laboratory).  This means that the null hypothesis that the effect is say non-positive is true in one laboratory, but false in the other laboratory, and thus the positive effect is not replicated  in both laboratories.
Replicability problems also emerge in medical research and are of great concern. Half of phase III clinical trials fail even though they rely on one of the many measures of success that were studied in the phase II trials (\cite{kaplan08} and \cite{benjamini13}). This suggests that the therapeutic effect discovered in the phase II study, which leads to the phase III study conducted in different patients, using similar but not identical methods and measures of success, was not replicated. Obviously the phase III studies were not under-powered, so failing to discover an effect suggests that the effect was absent from phase III even if it was present in phase II. %Another explanation may be that the effect was absent in phase II and the statistical analysis may have been inadequate.
%Not only in clinical trials YOAV reference of Colleague and Ioannidis 50 papers checked if repeated experiment.  In popular literature, New Yorker, Economist,  YOAV reference.
In genomic research, the interest is in the genetic effect on phenotype. In different studies of the same associations with phenotype, we seem to be testing the same hypotheses but the hypotheses tested are actually much more particular. Whether a hypothesis is true may depend
 on the cohorts in the study, that are from specific populations exposed to specific environments (for particular examples, see section Results). However,  if discoveries are made, it is of great interest to see whether these discoveries are replicated in different cohorts, from different populations, with different environmental exposures and different measurement techniques. The paramount importance of having replicated findings is well
recognized in genomic research \cite{Lander95}. In particular,
this is so in genome-wide association studies (GWAS), see
\cite{McCarthy08} and \cite{NCI07}. As noted in  \cite{kraft09}, the
anticipated effects for common variants in GWAS are modest and very
similar in magnitude to the subtle biases that may affect genetic
association studies - most notably population stratification bias.
For this reason, they argue that it is important to observe the same
association in other studies using similar, but not identical,
sub-populations and methods. Obviously, splitting the data into two independent parts and doing the same analysis on each does not answer the above concerns.

Replicability problems arise in many additional scientific areas, and discussions of these problems reached prominent general-interest venues, for instance the New Yorker (December 13, 2010) and the Economist (October 19, 2013).
We need to have an objective way to declare that a certain study really replicates the findings in another study. This paper makes a concrete, objective, easy to apply, and rigorously motivated way to determine that a finding has been replicated.

{\bf Replicability versus meta-analysis.} In many areas it is common to combine the results of studies that examine the same hypotheses by a meta-analysis.
Pooling results across studies is especially attractive when single studies are underpowered, utilizing the potential increase in power of combining the studies, but the meta-analysis $p$-value tests only the null hypothesis of no signal in all studies. As a result, a strong signal in one of the studies (with $p$-value close to zero)
is enough to declare the meta-analysis finding as highly significant.

A meta-analysis discovery based on a few studies is no better than a discovery from a single large study in assessing replicability, unless we are ready to assume that if a signal exists in one study it exists in all, i.e. that a discovery has to be a replicated discovery. This obviously need not be true, as discussed above. Similarly, replicability cannot be assessed in the following common practice: features are screened in a primary study, then the features with promising results are examined in a follow-up study, and the discoveries are only based on the results from the follow-up study. These are follow-up study discoveries, not discoveries that replicated from the primary study to the follow-up study.

In GWAS,
a typical table of results reports the
$p$-values in the primary and follow-up study, side by side, as well as the
meta-analysis $p$-values, for the SNPs with the smallest meta-analysis $p$-values.
Table 1 columns 1-6 is an example of such a table of results \cite{Yu12}.
In replicability analysis, the null hypothesis of signal in at most one study is tested, the rejection of which yields the statistical significance of the replicability claim. (Replicability is sometimes referred to as reproducibility, but see \cite{Peng09}.)

{\bf The $r$-value for replicability.} If a hypothesis in  one study is rejected at the 0.05 level, and it is also rejected in the same direction in another study at the 0.05 level, then replicability is intuitively established. This is also a sound statistical claim, in the sense that the probability of claiming that a finding is replicated if the null hypothesis is true in at least one of the studies is at most 0.05.
 The need for a statistical framework for establishing replicability becomes essential with the use of high throughput methods.  The potential to err in inference when more than one study is involved is more severe when each study is examining simultaneously many features. The choices for selection are much wider. Therefore, the statistical methods needed are more complicated than the very intuitive statistical method  for establishing replicability when a single feature is involved. %Even for a single study raw $p$-values are not useful enough,
%because few features are being selected from the many scanned.

Multiple testing methods are widely employed to adjust for the effect of selection,
either by controlling the probability of erroneously selecting even a single feature (FWER),
or by controlling the false discovery rate (FDR).
The concern regarding the selected claims of replicability is even greater,
because the selection takes place both after the primary study and after the follow-up study.
Our method reports the {\it r-value}, that can be defined for either error rates for replicability analysis. Here we emphasize the FDR: \\
{\noindent \bf Definition.}{\sl The FDR  $r$-value for feature $i$ is the lowest FDR  level  at which we can say that the finding is among the replicated ones.}\\
The smaller the $r$-value, the stronger the evidence in favor of  replicability of the finding.
It can be compared to any desired level of FDR in the same way that a $p$-value is commonly compared to the desired false detection parameter $\alpha$.

In this work we introduce a method for computing $r$-values for features examined in primary and follow-up studies. We  suggest to complement tables of results that report for selected findings the primary, follow-up, and meta-analysis $p$-values, with an additional column of $r$-values.
The $r$-values in column 7 of Table 1 are all below
0.05, concurring with the main replicability findings of \cite{Yu12}.
The ranking of $r$-values is different than the ranking of the
meta-analysis $p$-values, indicating the novelty of the added information.
Table 2 shows the results of a somewhat more complicated example to be discussed below,
where the difference between the meta-analysis and the replicability conclusions is more dramatic.

\section{Assessing replicability from follow-up studies}\label{sec-methods}
We will concentrate on the widely used design in ``omics" that  examines $m$  features in the primary study, and only a fraction thereof in the follow-up study.  For other designs, see the Section on ``Assessing replicability in other designs".

When $m=1$, as we discussed in the introduction, replicability is established at the 0.05 significance level if both $p$-values are at most 0.05. When $m>1$, this design can be analyzed by applying a multiple testing procedure on the maximum of the two studies $p$-values, setting conservatively the maximum value at one if the feature was not followed-up.
This is not recommended since the price paid for multiplicity is too large.  More powerful procedures for FWER and FDR control were suggested for this design in \cite{Bogomolov12}, in which effectively the primary study $p$-values have to be adjusted for the multiplicity of $m$ hypotheses, but the follow-up study $p$-values need to be adjusted only for the multiplicity of the hypotheses followed up.
Here we suggest a generalization of the method of \cite{Bogomolov12}, which offers further power gain in the typical situation in ``omics" research where most of the hypotheses examined in the primary study are true null hypotheses. We demonstrate our proposal on $p$-values from GWAS. However, the $p$-values can obviously come from other applications such as  exome-sequencing studies, ChIP experiments, or microarray studies.

Let $f_{00}$ denote the  fraction of features, out of the $m$ features examined in the primary study,  that are null in both studies. We cannot estimate $f_{00}$ from the data, since only a handful
 of promising features (SNPs) are followed up in practice. However, $f_{00}$ is typically closer to one than to
zero, and  we can give a conservative guess for a lower bound on $f_{00}$, call it $l_{00}$. In typical GWAS on the whole genome, $l_{00}=0.8$ is conservative. We can exploit the fact that $l_{00}>0$ to gain power.

\subsection{Computation of $r$-values for FDR-replicability}
\begin{enumerate}
\item Data input:
\begin{enumerate}
\item $m$, the number of features examined in the primary study.
\item $\mathcal R_1$, the set of features selected for follow-up based on primary study results. Let $R_1 = |\mathcal R_1|$ be their number.
\item $\{(p_{1j},p_{2j}): j\in\mathcal R_1\}$, where  $p_{1j}$ and $p_{2j}$ are, respectively, the primary and follow-up study $p$-values for feature  $j\in\mathcal R_1$.
\end{enumerate}
\item Parameters input:
\begin{enumerate}
\item $l_{00}\in [0,1)$, the lower bound on $f_{00}$ (see above), default value for whole genome GWAS is $l_{00}=0.8$.
\item $c_2\in (0,1)$, the emphasis given to the follow-up study (see Section Variations), default value is $c_2 =0.5$.
\end{enumerate}
\item Definition of the functions $f_i(x), i \in \mathcal R_1, x\in (0,1)$:
\begin{enumerate}
\item Compute $c_1 = \frac{1-c_2}{1-l_{00}(1-c_2x)}$.
\item For every feature $j\in \mathcal R_1 $ compute the following $e$-values
$$e_j=\max\left(\frac{1}{c_1}p_{1j},\,\frac{R_1}{m c_2}p_{2j}\right), j
\in \mathcal{R}_1.$$
\item Let $f_i(x) = \min_{\{j: e_j\geq e_i, j \in \mathcal R_1\}}\frac{e_j
m }{rank(e_j)}$, where $rank(e_j)$ is the rank of the $e$-value for feature $j\in \mathcal R_1$ (with maximum rank for ties).
\end{enumerate}
\item The FDR $r$-value for feature $i\in \mathcal R_1$ is the solution to $f_i(r_i) = r_i$ if a solution exists in $(0,1)$,  and 1 otherwise. The solution is unique, see SI Lemma S1.1 for a proof.
\end{enumerate}

The $r$-values can be computed using our web application \url{http://www.math.tau.ac.il/$\sim$ruheller/App.html}. An R script is also available in RunMyCode,\url{http://www.runmycode.org/companion/view/542}.

The adjustment in Step 3(c)  is similar to the computation of the adjusted $p$-values \cite{Reiner03} for the Benjamini-Hochberg (BH) procedure \cite{yoav1},
 the important
difference being that $e$-values are used instead of $p$-values.
The replicability claims at a prefixed level $q$, say $q=0.05$, are all indices with $r$-values at most $0.05$.  The FDR for replicability analysis is then controlled at level 0.05, see Section Derivation and Properties for details.

%For the choice
%$l_{00}=0.8$, which is a conservative lower bound for typical GWAS on the whole genome,   $c_1(q) \approx 2.5$ for $q$ small. This parameter choice
%results in $r$-values that are up to five times smaller than the
%$r$-values  with $l_{00} = 0$.
For $l_{00} = 0$, declaring as replicated  the findings with $r$-values at most $q$ coincides with Procedure 3.2 in
 \cite{Bogomolov12}. It is easy to see that with $l_{00}>0$, we will have at least as many replicability claims as with Procedure 3.2 in
 \cite{Bogomolov12}.
 %Procedure \ref{proc} with parameters $(l_{00},c_2,q)$, where $l_{00}>0$, will have at least as many replicability claims as Procedure 3.2 in
 %\cite{Bogomolov12}.
 Next we show in GWAS examples and simulations that the power increases with $l_{00}$, and can lead to many more discoveries than with Procedure 3.2 in
 \cite{Bogomolov12}, while maintaining FDR control.

\subsection{Results}\label{subsec-examples}
We consider three recent articles reporting GWAS, where
hundreds of thousands of SNPs are examined in the primary studies,
and only a small fraction  of these SNPs are examined in the
follow-up studies. In these examples, the primary and follow-up studies differ in the sub-populations examined, and may also differ in design and analysis.
In addition, the primary and follow-up studies may differ in quality. It is therefore of scientific importance to discover which associations were replicated.
%We quantify the evidence of replicability of
%associations for the SNPs followed-up in terms of the $r$-values for
%three values of the lower bound: $l_{00} \in \{0.8,0.5,0\}$. We used the approximation $c_1\approx \frac{1-c_2}{1-l_{00}}$, so there was no need to prefix $q$ for the $r$-value computation.
The examples differ in design, and in the selection rules for forwarding SNPs for follow-up. In the first example, there is one primary study and one follow-up study, few dozen SNPs are followed up, and only a handful have $r$-values below 0.05. In the second example, the primary study is a meta-analysis of three studies, more than a hundred hypotheses are followed-up, and few dozen SNPs have $r$-values below 0.05.  In the third example, there were three stages: a primary study, then a follow-up study, and then an additional follow-up study that was based on the first follow-up study.

Our first example is  GWAS of IgA nephropathy in Han Chinese \cite{Yu12}. To
discover association between SNPs and IgA nephropathy,
 444882 SNPs were genotyped
 in 1523 cases from southern China, and 4276 controls from Singapore and from southern and northern China, with the same ancestral
 origin. For follow-up, 61 SNPs were
 measured in two studies: 1402 cases and 1716 controls from northern
 China, and 1301 cases and 1748 controls from southern China.
 The 61 SNPs selected for follow-up had primary study $p$-values below
 $10^{-5}$. Table 1 shows the seven SNPs with the smallest
meta-analysis $p$-values, out of the $61$ SNPs followed up.
The associations for these seven SNPs have been replicated with $r$-values $\leq0.05$ for $l_{00}=0.8$.
 The seven SNPs clearly stand out from the
remaining 54 SNPs followed-up, that have $r$-values of one, see
Table S1 in the Supporting Information (SI). If the researcher is willing
to assume only a lower bound of 0.5 or of zero for $f_{00}$, then
  the $r$-values are larger than with $l_{00}=0.8$.
Table S1 in the SI shows that with $l_{00}=0.5$ and $l_{00} = 0$, respectively, only six and five SNPs had $r$-values below 0.05.
%six
%SNPs had $r$-values below 0.05, and with $l_{00}=0$, only five SNPs had
%$r$-values below 0.05.

Our second example is GWAS of Crohn's disease (CD). To discover
associations between SNPs and CD, \cite{Barrett08} examined 635547
SNPs on 3230 cases and 4829 controls of European descent,
  collected in three separate studies: NIDDK4, WTCCC5, and a Belgian-French study.  For follow-up,
  126 SNPs were measured in 2325 additional cases and 1809
  controls as well as in an independent family-based dataset of 1339
  trios of parents and their affected offspring. The two smallest $p$-values in each distinct region with  primary study $p$-values below $5\times 10^{-5}$ were considered for follow-up.  Table S2 in the SI shows the $126$ SNPs
followed-up. Applying our proposal with parameter $l_{00}=0.8$, we decide that 52 SNPs have replicated associations at $r$-values $\leq 0.05$.
 The 52 SNPs with
replicated associations did not correspond to the 52 SNPs with the
smallest meta-analysis $p$-values. For example, the SNP in row 35
had the 35th smallest meta-analysis $p$-value, but its $r$-value was
$0.09$, thus it was not among the 52  replicated discoveries.%\footnote{The SNPs with replicated associations were a subset of the SNPs associated with CD in at least one study, and most associations discovered were replicated associations. Specifics follow. In order to identify the SNPs with evidence of associations, while controlling the FDR,
%the following typical meta-analysis can be done: applying the BH
%procedure at level 0.05 on the combined data from the primary and
%follow-up studies, by using the meta-analysis $p$-values when
%follow-up information was available and primary study $p$-values
%when there is no follow-up information. Since we only had primary
%and follow-up $p$-values for the 126 SNPs selected for follow-up, we
%set the remaining 635421 $p$-values conservatively as one. We
%discovered 76 SNPs associated with the disease in at least one of
%the studies, and the 52 replicated discoveries were a subset
%thereof.}.
The last column of Table S2 in the SI
marks the 30 SNPs that were highlighted as ``convincingly
(Bonferroni $P <0.05$) replicated CD risk loci", based on the
follow-up study $p$-values, in Table 1 of the main manuscript of
\cite{Barrett08}. These 30 SNPs have $r$-values below 0.05, so they
are a subset of the 52 replicated discoveries. Our replicability
analysis discovers more loci, in particular three loci (rows 34, 44,
and 59 in Table S2 of the SI) that did not reach
the conservative Bonferroni threshold of \cite{Barrett08} on the
follow-up study $p$-values, yet were pointed out in Table 2 of
\cite{Barrett08} to be "Nominally (uncorrected $P<0.05$) replicated
CD risk loci".

Our third example is GWAS of type 2 diabetes (T2D). To discover
association between SNPs and T2D, \cite{Zeggini08} examined more
than two million  SNPs imputed from about 400000 SNPs collected on
4549 cases and 5579 controls combined from three separate studies:
DGI, WTCCC, and FUSION. For follow-up, 68 SNPs were measured in
10037 cases and 12389 controls combined from additional genotyping
of DGI, WTCCC, and FUSION. The 68 SNPs chosen for follow-up had
primary study $p$-values below $10^{-4}$, and they were in loci that
were not discovered in previous studies. For additional follow-up,
11  out of the 68 SNPs were measured in 14157 cases and 43209
controls of European descent combined from 10 centers. The 11 SNPs
forwarded for an additional follow-up had $p$-values below 0.005 in
the first follow-up study, as well as meta-analysis $p$-values
below $10^{-5}$ when combining the evidence from the primary study
and the first follow-up study. While there was no evidence of
replicability from the primary study to the follow-up studies, there
was evidence of replicability from the first follow-up study to the
second follow-up study. Table 2 shows the $11$
SNPs followed-up from the first follow-up study to the second
follow-up study. Applying our proposal with $l_{00}=0$, we decide that five SNPs have replicated associations with $r$-values $\leq 0.05$.
Note that we set $l_{00}=0$ since most of the 68 SNPs in the first follow-up
study are already believed to be associated with the disease.

\section{Derivation and properties} %EDITOR SUGGESTION: GATHER HERE ALL THE EARLIER STUFF THAT'S ABOUT ASIDES TO EXPERTS
Here we give the formal framework for replicability analysis, and the theoretical properties of our proposal.
The family of $m$ features examined in the primary study, indexed
by $I = \{ 1,\ldots,m\}$,  may be divided into four sub-families
with the following indices: $I_{00}$, $I_{01}$, $I_{10}$, and $I_{11}$,
for the features with hypotheses that are, respectively, null in both studies, null in the primary study only, null in the follow-up study only, and non-null in both studies.
%$I_{00}$, for the features with hypotheses
%that are null in both studies; $I_{01}$, for features with the
% hypotheses that are null in the primary study only;
%$I_{10}$, for features with the  hypotheses that are null in the
%follow-up study only; and $I_{11}$, for the features with the hypotheses
%that are non-null in both studies.
Suppose  $R$ replicability claims are made by an analysis. Denoting by $R_{ij}$ the number of replicability claims from sub-family $I_{ij}$, $R_{11}$ is the number of true replicability claims, and $R-R_{11} = R_{00}+R_{01}+R_{10}$ is the number of false replicability claims.

The FDR for replicability analysis is the expected proportion of false
replicability claims among all those called replicated:
$$FDR = E\left(\frac{R_{00}+R_{01}+R_{10}}{\max (R,1)}\right).$$

{\noindent \bf Definition. }{\sl  A stable selection rule
satisfies the following condition: for any $j\in \mathcal{R}_1$,
fixing all primary study $p$-values except for $p_{1j}$ and changing $p_{1j}$ so
that $j$ is still selected, will not change the set $\mathcal R_1$.}

% This is a technical condition that is typically satisfied.
Stable selection rules include
 selecting the hypotheses with $p$-values below a certain cut-off, or by a
 non-adaptive multiple testing procedure on the primary study $p$-values such as the BH procedure
 for FDR control or the Bonferroni procedure for FWER control, or selecting the $k$ hypotheses with the smallest
 $p$-values, where $k$ is fixed in advance. %In all the examples considered henceforth, the selection rule is stable.

\begin{theorem}\label{theorem-upperbound1}
A procedure that declares findings with $r$-values at most $q$ as replicated controls the FDR for replicability analysis at level  at most $q$ if
the rule by which the set $\mathcal{R}_1$ is selected is a stable selection rule, $l_{00}\leq f_{00}$, the $p$-values
within the follow-up study are jointly independent or are positive regression dependent on the subset of true null hypotheses (property
PRDS), and are independent of the primary study $p$-values, in either
one of the following situations:
\begin{enumerate}
\item The $p$-values within the primary study are
independent.
\item Arbitrary dependence among the $p$-values within the primary
study, when in Step 3  $m$ is replaced by
$m^*=m\sum_{i=1}^m1/i.$
\item Arbitrary dependence among the $p$-values within the primary study, and the selection rule is such that the primary study $p$-values of the features that are selected for follow-up
are at most a fixed threshold $t\in (0,1)$, when $c_1$  computed in Step 3(a) is replaced by
$$\tilde{c}_1(x)=\max\{a:\,a(1+\sum_{i=1}^{\lceil tm/(ax)-1\rceil}1/i)=c_1(x)\},$$
where $c_1(x) = \frac{1-c_2}{1-l_{00}(1-c_2x)}$. Steps 3(b) and 3(c) remain unchanged. In step 4, the FDR $r$-value for feature $i\in \mathcal R_1$  is $r_i= \min\{x: f_i(x)\leq x\}$ if a solution exists in $(0,1)$, and one otherwise.
\end{enumerate}
\end{theorem}
See the SI for a proof. The implication of item 3 is that for FDR-replicability at level $q$, if $t\leq c_1(q)q/m$, no modification is required, so the procedure that declares as replicated all features with $r$-values at most $q$ controls the FDR at level $q$ on replicability claims for any type of dependency in the primary study. Note that the modification in item 3 will lead to more discoveries than the modification in item 2 only if $t<\frac{c_1(q)q}{1+\sum_{i=1}^{m-1}1/i}$.

In the SI we  show realistic GWAS simulations that preserve the dependency across $p$-values in each study. For  $l_{00}\in \{0, 0.8,0.9,0.95,0.99 \}$, the FDR of the procedure that declares findings with $r$-values (computed in Steps 1-4 of the original proposal)  at most $0.05$ as replicated is controlled below level $0.05$, suggesting that this procedure is valid for the type of dependency that occurs in GWAS. Since this procedure can be viewed as a two dimensional variant of the BH procedure, and the BH procedure is known to be robust to many types of dependencies, we conjecture that for $l_{00}\leq f_{00}$, our procedure controls the FDR at the nominal level $q$ for most types of dependencies that occur in practice, even if hypotheses with primary study $p$-values above $c_1(q)q/m$ are followed-up. In Table S5 of the SI we further show the superior power of our procedure over applying the BH procedure on the maximum of the two studies $p$-values (at level $0.05/(1-l_{00})$, where the maximum value is set to one for $j\notin \mathcal R_1$).

\section{Variations}
%Here we discuss variations on our main proposal,  and an alternative proposal for FWER replicability analysis.
\subsection{Choice of emphasis between the studies}\label{subsec-c2choice}
The $e$-value computation requires combining the $p$-values from the primary and the follow-up study using a parameter $c_2$, which we set to be $c_2=0.5$ in the computation above. More generally, for FDR control we need to first select $c_2\in (0,1)$. We  shall show the effect the choice of $c_2$ has on  the $r$-values for given $p$-values, and  argue from power considerations that the choice $c_2=0.5$ is reasonable.

The following procedure is identical to that of declaring the set of findings with $r$-values  at most $q$ as replicated, see proof in SI Lemma S1.1. First, compute the number of replicability claims at level $q$ as follows:
 $$R_2\triangleq\max\left\{r:
\sum_{j\in\mathcal{R}_1}\textbf{I}\left[(p_{1j},
p_{2j})\leq\left(\frac{r}{m}c_1(q)q, \frac{r}{R_1}c_2 q\right)\right]
= r\right\}.$$
Next, declare as replicated findings the set
$$\mathcal R_2= \left\{j: (p_{1j}, p_{2j})\leq\left(\frac{R_2}{m}c_1(q)q,
\frac{R_2}{R_1}c_2q\right), j \in \mathcal R_1\right\}.$$

From this equivalent procedure it is clear that a larger choice  $c_2\in (0,1)$ will make the threshold that $p_{2j}$ has to pass larger, but the threshold that $p_{1j}$ has to pass smaller, so for the extreme choice   $c_2\approx 1$, the discovered findings can only be features with tiny primary study $p$-values, and for the extreme choice of $c_2 \approx 0$, the discovered findings can only be features with tiny follow-up study $p$-values. For $q$ small, the primary and follow-up study $p$-values will have the same threshold if $\frac{1}{m}\frac{(1-c_2)}{1-l_{00}}= \frac{c_2}{R_1}$, i.e. $c_2 = \frac{1}{1+m(1-l_{00})/R_1}$, which is close to zero if $R_1/m$ is very small (as is typical in GWAS). Therefore, this choice is not recommended unless the power of the follow-up study is extremely large. For the choice $c_2=0.5$, the threshold for the follow-up study $p$-value  is larger than for the primary study $p$-value by the factor  $m(1-l_{00})/R_1$, i.e. the ratio of the number of hypotheses that should be adjusted for in the primary study to that in the follow-up study. We show next that this choice is good from efficiency considerations.

In simulations, detailed in the
SI, we observed that for a given $l_{00}$  the
optimal $c_2$, i.e. the choice of $c_2$ that maximizes power, has only a
small gain in power over the choice $c_2 = 0.5$. We considered $m=1000$ SNPs, out of
which $f_{00}=0.9$ had no signal, $f_{01}=0.025$ had
signal only in the follow-up study, $f_{10}=0.025$ had signal
only in the primary study, and $f_{11}= 0.05$ had signal in both
studies. The power to detect the signal in the primary study was set
to be $\pi_1=0.1$ for a threshold of $0.05/m$, and the power to
detect the signal in the follow-up study was set to be $\pi_2\in
\{0.8, 0.5, 0.2 \}$ for a threshold of $0.05/R_1$. The selection
rule for follow-up  was the BH procedure at level $c_1(q)q$ on the primary study $p$-values, with
$q=0.05$. See Section S3 in the SI for a discussion of the advantage of this selection rule over other selection rules.

The power increased with $l_{00}$ as well as with $\pi_2$.
% Moreover, the
%difference in power between the best choice of
%$c_2$ (which is unknown in advance), and
%our suggestion $c_2 = 0.5$, is small.
In the SI, Table S4 shows that the gain in power of using $l_{00}>0$ over
$l_{00}=0$ can be large. Figure
S1 shows the average power and the power for at least one true
replicability discovery as a function of $c_2$. %In all simulations, in Step 1 of the $r$-value computation we used the approximation $c_1\approx \frac{1-c_2}{1-l_{00}}$.

Our simulations mimic the typical setting in GWAS on the whole
genome, where SNPs that are associated with the phenotype have
typically low power (0.1 in the above simulations) to pass the
severe Bonferroni threshold of the large number of hypotheses
examined in the primary study, yet the power to pass the far less
severe Bonferroni threshold of the few dozen hypotheses examined in
the follow-up study is greater (0.2, 0.5, or 0.8 in the above
simulations). Therefore, for GWAS on the whole genome, we recommend
setting $c_2=0.5$.

\subsection{FWER-replicability}\label{subsec-FWER}
 The FWER
criterion, $$FWER = \textmd{Pr}(R_{00}+R_{01}+R_{10}>0),$$ is more
stringent than the FDR, yet it may sometimes be desired. % be preferred over the FDR controlling procedure suggested above when very few hypotheses are followed-up.
We define the FWER $r$-value as the lowest FWER level at which we can say that the finding has been significantly replicated. The $r$-value can be compared to any desired level of FWER.
An FWER controlling procedure for replicability analysis was
suggested in \cite{Bogomolov12}: it applies an FWER controlling
procedure at level $c_1\alpha$ on the primary study $p$-values, and
at level $c_2\alpha$ on the subset of discoveries from the primary
study that were followed-up, where $c_1+c_2 = 1$.
If a non-zero lower bound on $f_{00}$ is available,
then this lower bound can be used in order to  choose parameters
$(c_1,c_2)$ with a sum greater than one. Specifically, for FWER
control using Bonferroni, the data input and parameters input is the same as in our proposal for FDR-replicability in Steps 1 and 2, but the computation in Step 3 is different.
For feature $j\in \mathcal{R}_1$,
$$f^{Bonf}_j(x) = \max \left(mp_{1j}/c_1,
|\mathcal{R}_1|p_{2j}/c_2\right), \quad c_1 = \frac{1-c_2}{1-l_{00}(1-c_2x)}.$$

The Bonferroni $r$-value for feature $j$ is the solution to $f^{Bonf}_j(r_j) = r_j$ if a solution exists in $[0,1)$, and one otherwise. The replicability claims at a prefixed level $\alpha$, say
$\alpha = 0.05$, are all indices with $r$-values at most 0.05. The FWER for replicability analysis is then controlled at level 0.05, see
SI for the proof.

%For example, with $c_2=0.5$,
%the discovery is considered replicated if the primary
%study $p$-value is at most $0.5\alpha/[(1-l_{00})m]$ and the
%follow-up $p$-value is at most $0.5\alpha/R_1$.

We computed the Bonferroni $r$-values in a GWAS of thyrotoxic periodic
paralysis (TPP) \cite{Cheung12}. In 70 cases and 800 controls from the Hong Kong (Southern) Chinese population,  486782 SNPs were genotyped.
Table S3 shows the four most significant SNPs followed-up in additional 54
 southern Chinese TPP cases and 400 healthy Taiwanese controls. The associations were successfully replicated with
 Bonferroni $r$-values far below 0.05, concurring with the
claim in \cite{Cheung12} that ``Associations for all four SNPs were
successfully replicated".

\section{Assessing replicability in other designs}
The concept of the $r$-value is also relevant to the communication of the results of replicability in other designs.
If $n>2$ studies examine a single feature, then replicability of the finding in all $n$ studies is established at the 0.05 significance level if the maximum  $p$-value is at most 0.05.
However, if a weaker notion of replicability is of interest, e.g. that the finding has been replicated in at least two studies, then the evidence towards replicability can be computed as follows. First, for every subset of $n-1$ studies, a meta-analysis $p$-value is computed. Then, replicability in at least two studies is established at the 0.05 significance level if the maximum of the $n$ meta-analysis $p$-values is at most 0.05. This can be generalized to discover whether the finding has been replicated in at least $u$ studies, where $u \in \{2,\ldots, n \}$, as detailed in \cite{conj}.
%More generally, if a finding is defined to be replicated across studies if the hypothesis is non-null in at least $u$ studes, where $u \in \{2,\ldots, n \}$, then first for every subset of $n-u+1$ studies, a
%meta-analysis $p$-value is computed, and then, replicability is established at the 0.05 significance level if the maximum of the $\binom{n}{n-u+1}$ meta-analysis $p$-values is at most 0.05. See \cite{conj} for more details.

If $n\geq 2$ studies examine each $m>1$ features, then for each $i\in \{1,\ldots,m\}$ the  $p$-value for testing for replicability can be computed as above, but instead of comparing each  to 0.05, the BH procedure  is applied and the discoveries are considered as replicated findings.  The procedure was suggested in \cite{benjamini09}, and for $n=2$ it amounts to using the maximum of the two studies $p$-values for each feature in the BH procedure. The power of this procedure may be low when a large fraction of the null hypotheses are true, since the
null hypothesis for replicability analysis is not simple, and the BH procedure is applied on a
set of $p$-values that may have a null distribution that is
stochastically much larger than uniform.  The loss of power of multiple testing procedures can indeed be severe when using
over-conservative $p$-values from composite null hypotheses \cite{Dickhaus13}. An empirical Bayes approach for discovering whether results have been replicated across studies was suggested in  \cite{Heller13}, and compared
with the analysis of \cite{benjamini09}, concluding that the empirical Bayes analysis discovers many
more replicated findings. The accuracy of the
empirical Bayes analysis relies on the ability to estimate well the unknown parameters, and thus it is suitable  in problems such as GWAS, where each study
contains hundreds of thousands of SNPs, and the dependency across SNPs is local, but may not be suitable for applications with a smaller number of features and non-local dependency.  A method based on relative ranking of the $p$-values to control their ``irreproducible discovery rate" was suggested in \cite{Li11}. A list-intersection test to compare top ranked gene lists from multiple studies to discover the common significant set of genes was suggested in \cite{Natarajan12}.

To summarize, although for $m=1$ there is a straightforward solution for the problem of establishing replicability, once we move away from this simple setting the problem is more complicated. For designs with more than one potential finding, it is very useful to quantify and report the evidence towards replicability by an $r$-value. The $r$-value is a general concept, but the $r$-value computation depends on the multiple testing procedure used, which in turn depends on the design of the replicability problem.

\section{Discussion}\label{sec-discussion}
 The $r$-value was coined in the FDR context, in accordance with the commonly used $q$-value \cite{Storey02b}. We proposed the  $r$-value as an FDR-based measure of significance
for replicability analysis. We showed in GWAS examples that
the smallest meta-analysis $p$-values may not have the strongest
evidence towards replicability of association, and we
suggest to report the $r$-values in addition to the meta-analysis
$p$-values in the table of results.

  In practice, the primary study $p$-values are rarely independent. We prove that our main proposal controls the FDR on replicability claims if the primary study $p$-values are independent, and suggest modifications of the proposal that are more conservative but have the theoretical guarantee of FDR control for any type of dependency among the primary study $p$-values. From empirical investigations, we conjecture that the conservative modifications in items 2 and 3 of Theorem 1 are unnecessary for the types of dependencies encountered in GWAS.  For our second example, of GWAS in CD, applying the more conservative proposal in item 2 of Theorem 1 resulted in 34 discoveries.
 In future research we plan to investigate theoretically the effects of local dependency and positive dependency in the primary study.

%We have shown that it is also useful for FWER control. The specification of which procedure is applied
%should be clearly stated when reporting $r$-values. The FDR may be preferred over the FWER  if
%it is enough to guarantee that the expected fraction of false
%replicability claims among the replicability claims is small. %When
%only few  hypotheses are followed-up, replicability analysis with
%FWER control may be applied.

%We showed GWAS examples on the full genome, where the choice
%$l_{00}=0.8$ is very reasonable, as well as an example where only
%few dozen SNPs comprise the primary study of the replicability
%analysis, where the choice $l_{00}=0$ was more reasonable. In the
%latter example, it did not matter that the few dozen SNPs from the
%primary study were selected from few million SNPs in a previous
%study. However, since the few dozen SNPs were already believed to
%contain signal, the lower bound on the number of SNPs not associated
%with the phenotype in both studies should be set as zero. We showed
%in simulations that the gain in exploiting the a-priori knowledge
%that $f_{00}>l_{00}$, for $l_{00}\geq 0.5$, over setting
%$l_{00}=0$, can be large,  and therefore we recommend setting $l_{00}\geq 0.5$ when
%appropriate.

We saw examples where the primary study was comprised of more than
one study, and more than one follow-up study was
performed. In the current work, we used all the information from the
primary studies for selection for follow-up, and  to
establish replicability the meta-analysis $p$-values of the primary
studies and the meta-analysis $p$-values of the follow-up studies
were used. Alternative ways of combining the
evidence, that can also point to the pair of studies in which the
evidence of replication is strongest, will be considered in the
future. The scientific evidence of two out of
two (2/2) studies is more convincing than that of two out of three
(2/3) studies or two out of $n$ ($2/n$) studies, and
the scientific evidence of $3/n$ studies is more convincing than
that of $2/n$ towards replicability. In the future, we plan to
develop methods for computing the $r_{u/n}$-value, that quantifies
the evidence that the finding has been replicated in at least $u$
out of $n$ studies, for $2\leq u\leq n$. This problem has been
addressed in \cite{conj}, but as was shown in \cite{Bogomolov12}
alternatives along the lines of the procedures suggested here may
benefit from increased power.

A referee pointed out that  follow-up studies may be designed to give  more trustworthy data, using more expensive equipment, e.g. using PCR or fine linkage analysis. If the aim is to detect associations in the follow-up study, then there is no need to combine the evidence from the primary study with that of the follow-up study. However, if the aim is to detect replicated associations,  then it may be of interest to have  unequal penalties for the error of discovering a finding that is only true in the primary study, and the error of discovering a finding that is only true in the follow-up study. Developing procedures that give unequal penalties to  these two errors is a challenging and interesting problem for future research, which may be approached by utilizing weights \cite{Benjamini97}.

\paragraph{Acknowledgments}
The research of Yoav Benjamini was supported by ERC grant (PSARPS). The research of Ruth Heller was supported by the Israel Science Foundation (ISF)
Grant no. 2012896. We thank Shay Yaacoby for help in developing the web application, and Neri Kafkafi for useful discussions on the history of replicability in science. We thank the editor and the reviewers for the extensive comments that led to a substantial improvement of the manuscript.

%% PNAS does not support submission of supporting .tex files such as BibTeX.
%% Instead all references must be included in the article .tex document.
%% If you currently use BibTeX, your bibliography is formed because the
%% command \verb+\bibliography{}+ brings the <filename>.bbl file into your
%% .tex document. To conform to PNAS requirements, copy the reference listings
%% from your .bbl file and add them to the article .tex file, using the
%% bibliography environment described above.

%%  Contact pnas@nas.edu if you need assistance with your
%%  bibliography.

% Sample bibliography item in PNAS format:
%% \bibitem{in-text reference} comma-separated author names up to 5,
%% for more than 5 authors use first author last name et al. (year published)
%% article title  {\it Journal Name} volume #: start page-end page.
%% ie,
% \bibitem{Neuhaus} Neuhaus J-M, Sitcher L, Meins F, Jr, Boller T (1991)
% A short C-terminal sequence is necessary and sufficient for the
% targeting of chitinases to the plant vacuole.
% {\it Proc Natl Acad Sci USA} 88:10362-10366.

%% Enter the largest bibliography number in the facing curly brackets
%% following \begin{thebibliography}

% latex table generated in R 2.15.2 by xtable 1.7-0 package
% Thu Jan 03 17:02:45 2013
\begin{table}[!hbp]
\begin{center}
\caption{ Replicability analysis for FDR control for the study of
\cite{Yu12}: GWAS of IgA nephropathy in Han Chinese. The number of
SNPs in the primary study was $444882$, and 61 were followed-up.
For the seven most significant
meta-analysis $p$-values: the position (columns 1-3), the primary and
follow-up study $p$-values (column 4 and 5), the meta-analysis
$p$-values (column 6), and the $r$-values (column 7). See Table S1 of
the SI for the results for all 61 SNPs
followed-up. The lower bound for $f_{00}$ was $l_{00}=0.8$ for the $r$-value computation.}\label{tab-ex1main}
\begin{tabular}{lllrrrr}
  \hline
   Chr.  & Position & Gene & p1 & p2 &  p\_meta & $r$-value  \\
  \hline
 6 & 32685358 & HLA-DRB1 & 8.19e-08 & 8.57e-14 & 4.13e-20 & 0.0074  \\
 8 & 6810195 & DEFAs & 2.04e-07 & 1.25e-07 & 3.18e-14  & 0.0090 \\
 6 & 32779226 & HLA-DQA/B & 3.28e-08 & 3.57e-06 & 3.43e-13  & 0.0059 \\
 22 & 28753460 & MTMR3 & 2.30e-07 & 2.02e-05 & 1.17e-11  & 0.0090 \\
 6 & 30049922 & HLA-A & 4.05e-09 & 3.68e-04 & 1.74e-11  & 0.0090  \\
 17 & 7403693 & TNFSF13 & 1.50e-06 & 2.52e-05 & 9.40e-11  & 0.0413 \\
 17 & 7431901 & MPDU1 & 5.52e-07 & 3.16e-04 & 4.31e-10  & 0.0169 \\
   \hline
\end{tabular}
\end{center}
\end{table}

\begin{table}[!hbp]
\begin{center}
\caption{Replicability analysis for FDR control for the study of \cite{Zeggini08} on GWAS of T2D. The
number of SNPs in the first follow-up study was 68, and 11 were
followed-up to the second follow-up study. For these 11  SNPs: the positions (columns 1-2), the primary study $p$-values and first and second follow-up studies $p$-values (columns 3-5), the meta-analysis $p$-values from all 3 studies (column 6), and the $r$-values quantifying the evidence of replicability from  the first to the second follow-up study (column 7).
The lower bound for $f_{00}$ was $l_{00}=0$ for the $r$-value computation, since the set of SNPs in the first follow-up study are already believed to be associated with T2D. }\label{tab-ex3main}
\begin{tabular}{lllrrrr}
  \hline
  Chr.  & Position &p.primary & p1 & p2 & p\_meta & $r$-value   \\
  \hline
 7 & 27953796 & 1.55e-04 & 8.07e-05 & 1.34e-07 &  4.96e-14 & 0.0055  \\
 10 & 12368016 & 4.21e-04 & 5.40e-05 & 1.49e-04 &   1.21e-10 & 0.0055 \\
 12 & 69949369 & 1.80e-05 & 9.83e-03 & 4.35e-05 &   1.11e-09 & 0.1490 \\
 2 & 43644474 & 1.83e-04 & 1.62e-03 & 9.22e-05 &   1.12e-09 & 0.0441 \\
 3 & 64686944 & 5.44e-04 & 1.02e-04 & 3.47e-03 &   1.17e-08 & 0.0254 \\
 1 & 120230001 & 1.14e-04 & 2.89e-03 & 1.95e-03 &  4.10e-08 & 0.0604\\
 12 & 53385263 & 3.18e-05 & 3.11e-03 & 8.81e-03 &  1.79e-07 & 0.0604  \\
 3 & 12252845 & 1.05e-05 & 4.50e-03 & 1.22e-02 &  1.97e-07 & 0.0765  \\
 1 & 120149926 & 1.35e-03 & 1.17e-03 & 7.84e-03 &  4.04e-07 &  0.0431 \\
 6 & 43919740 & 5.41e-05 & 1.46e-03 & 9.49e-02 &  4.03e-06 &  0.2090\\
 2 & 60581582 & 3.38e-05 & 1.38e-03 & 6.54e-01 &  1.02e-04 & 1.0000 \\
   \hline
   \end{tabular}
\end{center}
\end{table}

\renewcommand{\thesection}{S\arabic{section}}
\renewcommand{\thetable}{S\arabic{table}}
\renewcommand{\thefigure}{S\arabic{figure}}
\setcounter{equation}{0} \setcounter{table}{0}
\setcounter{figure}{0}

\appendix
\section{Proof of Theorem 1}
The procedure that declares as replicated all features with
$r$-values $\leq q$ is equivalent to the procedure in
 Section Variations, where the choice of emphasis between the studies is discussed
(bottom of page 4), as proved in  Lemma \ref{lemmathm1}.
 We shall show that our proposal, in its most general
form (i.e. with $c_2\in (0,1)$), controls the FDR at level at most
\begin{equation}\label{eq-FDRbound}
f_{00}c_1(q)c_2q^2+f_{01}c_1(q)q+E\left(\frac{|I_{10}\cap
\mathcal{R}_1|}{\max(|\mathcal{R}_1|, 1)}\right)c_2q
\end{equation}
under the conditions of Theorem 1, where $f_{0j} = \frac{|I_{0j}|}m,
j\in \{0,1\}$, and $f_{10} = \frac{|I_{10}|}m$.

%The two last terms of the upper bound
% in equation [\ref{eq-FDRbound}] are reached when $p_{2j}\approx 0$ for $j\in I_{01}$ and $p_{1j}\approx 0$ for $j\in I_{10}$.
%Before proving the above inequality, we shall show why if the above
%inequality  holds and $l_{00}\leq f_{00}$, Theorem 1 follows.
Before proving the above upper bound on FDR, we shall show that if
the above upper bound holds and $l_{00}\leq f_{00}$, Theorem 1
follows. Note that if the constants $(l_{00},c_2)$ satisfy the
inequality
$$f_{00}c_1(q)c_2q + f_{01}c_1(q) +c_2\leq 1, $$ then the FDR for replicability analysis
is controlled at level at most $q$. This inequality holds for  any
choice of $(l_{00},c_2)$ that  satisfies the relationship
$$l_{00} \leq \frac{1-f_{01}-f_{00}c_2q}{1-c_2q}.
$$
Unfortunately,  $f_{00}$ and $f_{01}$ are not known. If the guess
for $l_{00}$ is indeed conservative, i.e. $l_{00}\leq f_{00}$, then
the above inequality holds since $f_{00} \leq 1-f_{01}$. Thus, for
any value $l_{00}\leq f_{00}$ and $c_2\in (0,1)$, the FDR for
replicability analysis is controlled at level at most $q$. %and
%Theorem 1 is complete.
%However, a  conservative lower bound for $f_{00}$,  $l_{00}\leq f_{00}$, may be available.  Since $f_{01} \leq 1-f_{00}$, for any value $l_{00}\leq f_{00}$, it follows that the FDR is controlled at level $q$ since
%$$l_{00} \leq \frac{1-f_{01}-f_{00}c_2q}{1-c_2q}.
%$$

\textbf{Proof for the upper bound in (\ref{eq-FDRbound}).} Let $R_j$
be the indicator of whether $j$ was declared replicated for
$j=1,\ldots,m$, and $R = \sum_{j=1}^m R_j$. The FDR for
replicability analysis is
\begin{equation}
 FDR=
E\left(\sum_{j\in
I_{00}}\frac{R_j}{\max(R,1)}\right)+E\left(\sum_{j\in
I_{01}}\frac{R_j}{\max(R,1)}\right)+E\left(\sum_{j\in
I_{10}}\frac{R_j}{\max(R,1)}\right).\label{fdr}
\end{equation}
For items 1-3 we shall find an upper bound for each of the terms,
specifically we shall show the following inequalities
(\ref{fin01})-(\ref{end00}).
\begin{align}E\left(\sum_{j\in I_{01}}\frac{R_j}{\max(R,1)}\right)\leq|I_{01}|\frac{c_1(q)q}{m}=f_{01}c_1(q)q,\label{fin01}\end{align}
\begin{align}E\left(\sum_{j\in I_{10}}\frac{R_j}{\max(R,1)}\right)\leq E\left(\frac{|I_{10}\cap
\mathcal{R}_1|}{\max(|\mathcal{R}_1|, 1)}\right)c_2q,
\label{fin10}\end{align}
\begin{align}
E\left(\sum_{j\in I_{00}}\frac{R_j}{\max(R,1)}\right) \leq
f_{00}c_1(q)c_2q^2.\label{end00}
\end{align}
Obviously the upper bounds in (\ref{fin01})-(\ref{end00}) and the
equality in (\ref{fdr}) complete the proof for the upper bound in
(\ref{eq-FDRbound}). The upper bounds in (\ref{fin01}) and
(\ref{fin10}) follow directly from \cite{Bogomolov12}. The key
difference from \cite{Bogomolov12} is the fact that we consider a
tighter upper bound for $E\left(\sum_{j\in I_{00}}
R_j/\max(R,1)\right)$ given in (\ref{end00}). We shall proceed to
prove inequality (\ref{end00}) for items 1-3. %thus completing the
%proof.

We shall start with the proof of item 1 for the case where the
$p$-values within the follow-up study are jointly independent.
Inequality (\ref{fin01}) follows from the derivations leading to
(A.3) in \cite{Bogomolov12}\footnote{Replacing $q_1$ with $c_1q,$
$q-q_1$ with $c_2q,$ and $|I_0|$ with $|I_{01}|$ in the derivations
leading to (A.3) in \cite{Bogomolov12} we obtain the proof of
inequality (\ref{fin01}). This replacement should be made in all the
derivations in \cite{Bogomolov12} used in this proof.}. Inequality
(\ref{fin10}) follows from the derivations leading to (A.7) in
\cite{Bogomolov12}, and by taking the expectation of the expression
in (A.7) over the primary study $p$-values.
%and using similar derivations to those leading to (A.7) in
%\cite{Bogomolov12}, we obtain for any $p_1=(p_{11},\ldots, p_{1m})$
%\begin{align*}E\left(\sum_{j\in I_{10}}\frac{R_j}{\max(R,1)}|P_1=p_1\right)\leq \frac{|I_{10}\cap
%\mathcal{R}_1(p_1)|}{\max(|\mathcal{R}_1(p_1)|, 1)}c_2q.\end{align*}
%It follows that
%\begin{align}E\left(\sum_{j\in I_{10}}\frac{R_j}{\max(R,1)}\right)\leq E\left(\frac{|I_{10}\cap
%\mathcal{R}_1|}{\max(|\mathcal{R}_1|,
%1)}\right)c_2q.\label{fin10}\end{align}
We shall now prove inequality (\ref{end00}). We recall the following
definitions from \cite{Bogomolov12}. Let $P_1^{(j)}$ and $P_2^{(j)}$
denote the vectors $P_1=(P_{11}, \ldots, P_{1m})$ and $P_2=(P_{21},
\ldots, P_{2m})$ with, respectively, $P_{1j}$ and $P_{2j}$ excluded.
For $j\in\{1,\ldots,m\}$ arbitrary fixed, let
$\mathcal{R}_1^{(j)}(P_1^{(j)})\subseteq\{1,\ldots,j-1,
j+1,\ldots,m\}$ be the subset of indices selected along with index
$j.$ Note that since the selection rule is stable, this subset is
fixed as long as $P_{1j}$ is such that $j$ is selected based on
$(P_1^{(j)}, P_{1j})$. For any $j\in\{1,\ldots,m\}$ and given
$P_1^{(j)}$, for $i\in\{1,\ldots,j-1,j+1,\ldots,m\}$
\begin{align*}
e_i^{(j)} = \left\{
\begin{array}{cl}
\max\left(\frac{P_{1i}}{c_1},\,\frac{(|\mathcal{R}_1^{(j)}(P_1^{(j)})|+1)P_{2i}}{mc_2}\right) & \text{if } i\in \mathcal{R}_1^{(j)}(P_1^{(j)}),\\
 \infty& \text{otherwise. } \\
\end{array} \right.
\end{align*}
Let $e^{(j)}_{(1)}\leq\ldots\leq e^{(j)}_{(m-1)}$ be the sorted
$e_i^{(j)}$s, and $e^{(j)}_{(0)}=0.$\footnote{The $e$-values are
closely related to $T$-values defined in Appendix A of
\cite{Bogomolov12}. Specifically, $e_i^{(j)}=T_iq/m$ for
$j\in\{1,\ldots,m\}$ and $i\in\{1,\ldots,j-1,j+1,\ldots,m\}.$} For
$r=1,\ldots,m$, we define $C_r^{(j)}$ as the event in which if $j\in
I_{00}\cup I_{01}\cup I_{10}$ is declared replicated, $r$ hypotheses
are declared replicated including $j$, which amounts to:
\begin{eqnarray}
 && C_r^{(j)}=  \{(P_1^{(j)},
P_2^{(j)}):  \,e^{(j)}_{(r-1)}\leq \frac{r q}{m},
e^{(j)}_{(r)}>\frac{(r+1)q}{m},
e^{(j)}_{(r+1)}>\frac{(r+2)q}{m},\ldots, e^{(j)}_{(m-1)}>q\}.
\nonumber
\end{eqnarray}
Note that given $P_1,$ for $r>|\mathcal{R}_1|$, $C_r^{(j)} =
\emptyset$, since exactly $|\mathcal{R}_1|-1$ $e^{(j)}_i$'s are
finite. Obviously, $C_r^{(j)}$ and $C_{r'}^{(j)}$ are disjoint
events for any $r\neq r',$ and $\cup_{r=1}^m C_r^{(j)}$ is the
entire space of $(P_1^{(j)}, P_2^{(j)})$. Therefore,
$\sum_{r=1}^m\textmd{Pr}\left(C_r^{(j)}\right)=1.$

Note that from the equivalent procedure in Section Variations the following equality follows.
\begin{align}
E\left(\sum_{j\in I_{00}}\frac{R_j}{\max(R,1)}\right) &= \sum_{j\in
I_{00}}\sum_{r=1}^m \frac 1r \textmd{Pr}\left(j \in \mathcal{R}_1,
P_{1j}\leq \frac{rc_1(q)q}{m}, P_{2j}\leq
\frac{rc_2q}{\max(|\mathcal{R}_1|, 1)},
C_r^{(j)}\right)\notag\\&\leq \sum_{j\in I_{00}}\sum_{r=1}^m \frac
1r \textmd{Pr}\left( P_{1j}\leq \frac{rc_1(q)q}{m}, P_{2j}\leq c_2q,
C_r^{(j)}\right)\label{indepc1}\\&\leq
c_2q\frac{c_1(q)q}{m}\sum_{j\in
I_{00}}\sum_{r=1}^m\textmd{Pr}(C_r^{(j)})=|I_{00}|c_2q\frac{c_1(q)q}{m}=f_{00}c_1(q)c_2q^2,\label{endzero}
\end{align}
where the inequality in (\ref{indepc1}) follows from the fact that
for any given realization  of $|\mathcal{R}_1|$ and value of $r$
such that
 $r>|\mathcal{R}_1|$, $C_r^{(j)} =
\emptyset,$ the inequality in (\ref{endzero}) follows from the
independence of the $p$-values and the fact that $P_{1j}$ and
$P_{2j}$ are null-hypothesis $p$-values, and the first equality in
(\ref{endzero}) follows from the fact that
$\sum_{r=1}^m\textmd{Pr}\left(C_r^{(j)}\right)=1,$
 thus
completing the proof of item 1 for the case where the $p$-values
within the follow-up study are independent.

We shall now prove item 1 for the case where the $p$-values within
the follow-up study have property PRDS. The inequalities
(\ref{fin01}) and (\ref{fin10}) for this case follow from the
results in the Supplementary Material of \cite{Bogomolov12}.
Specifically, inequality (\ref{fin01}) follows from the proof of
Theorem S3.1 in  \cite{Bogomolov12} and inequality (\ref{fin10}) follows
from the proof of item 2 in Lemma S2.1 in \cite{Bogomolov12}. %It follows from the proof of
%Theorem S3.1 in the Supplementary Material of \cite{Bogomolov12}
%that the inequality in (\ref{fin01}) holds in this case as well. In
%addition, it follows from the proof of item 2 in Lemma S2.1 in the
%Supplementary Material of \cite{Bogomolov12} that the inequality in
%(\ref{fin10})
%Note that the upper bound
%%for
%%for equation (\ref{eq1}) for $j\in I_{01}$
%in (\ref{fin01}) is derived only using the independence of the
%$p$-values within the primary study and independence of the
%$p$-values across the studies, therefore this upper bound holds when
%the set of $p$-values within the follow-up study has property PRDS.
%Hence, for $j\in I_{01}$ an upper bound for expression (\ref{eq1})
%is $c_1q/m.$
For $j\in I_{00}$ and an arbitrary fixed $p_1=(p_{11},\ldots,
p_{1m})$ such that $|\mathcal{R}_1(p_1)|>0,$
\begin{eqnarray}
&&E\left(\frac{R_j}{\max(R,1)}|P_1=p_1\right) = \nonumber \\
&& \sum_{r=1}^{|\mathcal{R}_1(p_1)|} \frac {I\left(j \in
\mathcal{R}_1(p_1), p_{1j}\leq \frac{rc_1(q)q}{m}\right)}{r}
\textmd{Pr}\left(P_{2j}\leq \frac{rc_2q}{|\mathcal{R}_1(p_1)|},
C_r^{(j)}|P_1=p_1\right) \nonumber \\
&& \leq  %\sum_{r=1}^{|\mathcal{R}_1(p_1)|} \frac 1r
%\textmd{Pr}\left(P_{1j}\leq \frac{|\mathcal{R}_1(p_1)|c_1q}{m},
%P_{2j}\leq \frac{rc_2q}{|\mathcal{R}_1(p_1)|},
%C_r^{(j)}|P_1=p_1\right) \nonumber
%\\&&=
 I\left(p_{1j}\leq \frac{|\mathcal{R}_1(p_1)|c_1(q)q}{m}, j\in \mathcal{R}_1(p_1)
\right)\sum_{r=1}^{|\mathcal{R}_1(p_1)|} \frac 1r \textmd{Pr}\left(
P_{2j}\leq \frac{rc_2q}{|\mathcal{R}_1(p_1)|},
C_r^{(j)}|P_1=p_1\right) \nonumber \\
&& =  I\left(p_{1j}\leq \frac{|\mathcal{R}_1(p_1)|c_1(q)q}{m}, j\in
\mathcal{R}_1(p_1) \right) \sum_{r=1}^{|\mathcal{R}_1(p_1)|} \frac
1r \textmd{Pr}\left( C_r^{(j)}|P_{2j}\leq
\frac{rc_2q}{|\mathcal{R}_1(p_1)|}, P_1=p_1 \right)\nonumber \\&&
\quad \quad \times\textmd{Pr}\left(
P_{2j}\leq \frac{rc_2q}{|\mathcal{R}_1(p_1)|}|P_1=p_1\right) \nonumber \\
&& \leq \frac{c_2q}{|\mathcal{R}_1(p_1)|}I\left(p_{1j}\leq
\frac{|\mathcal{R}_1(p_1)|c_1(q)q}{m}, j\in \mathcal{R}_1(p_1)
\right)\label{nullpv1} \\&& \quad \quad
\times\sum_{r=1}^{|\mathcal{R}_1(p_1)|} \textmd{Pr}\left(
C_r^{(j)}|P_{2j}\leq \frac{rc_2q}{|\mathcal{R}_1(p_1)|}, P_1=p_1
\right)\nonumber\\
&& \leq \frac{c_2q}{|\mathcal{R}_1(p_1)|}I\left(p_{1j}\leq
\frac{|\mathcal{R}_1(p_1)|c_1(q)q}{m}, j\in \mathcal{R}_1(p_1)
\right), \label{fincond1}
\end{eqnarray}
where inequality (\ref{nullpv1}) follows from the independence of
the $p$-values across the studies and the fact that $P_{2j}$ is a
null-hypothesis $p$-value. We shall now show that inequality
(\ref{fincond1}) holds. It follows from item 1 of Lemma S2.1 in the
Supplementary Material of \cite{Bogomolov12} that
$$\sum_{r=1}^{|\mathcal{R}_1(p_1)|} \textmd{Pr}\left(
C_r^{(j)}|P_{2j}\leq \frac{rc_2q}{|\mathcal{R}_1(p_1)|}, P_1=p_1
\right)\leq 1$$ for any $p_1=(p_{11},\ldots, p_{1m})$ and $j\in
I_{10}\cap \mathcal{R}_1(p_1).$ It is straightforward to verify that
this result holds for $j\in I_{00}\cap \mathcal{R}_1(p_1)$ as well,
yielding inequality
(\ref{fincond1}). %Using (\ref{fincond1}) we obtain %the upper bounds
%on expression (\ref{eq1}) for $j\in I_{00}$ and for $j\in I_{10}$.
%For $j\in I_{10}$, it follows that
%\begin{eqnarray*}
%&&E\left(\frac{R_j}{\max(R,1)}|P_1=p_1\right)
%\leq\frac{c_2q}{|\mathcal{R}_1(p_1)|} I\left(j \in
%\mathcal{R}_1(p_1)\right),
%\end{eqnarray*}
%therefore
%\begin{eqnarray}
%&&E\left(\frac{R_j}{\max(R,1)}\right) \leq  c_2qE\left(\frac{
%I\left(j \in \mathcal{R}_1\right)}{\max(|\mathcal{R}_1|,
%1)}\right).\label{fin10prds}
%\end{eqnarray}
It follows that for $j\in I_{00}$,
\begin{eqnarray}E\left(\frac{R_j}{\max(R,1)}\right)\leq
c_2qE\left[\frac{I\left(P_{1j}\leq
\frac{|\mathcal{R}_1(P_1)|c_1(q)q}{m}, j\in \mathcal{R}_1(P_1)
\right)}{\max(|\mathcal{R}_1(P_1)|, 1)} \right].\label{uncond1}
\end{eqnarray}
Note that for $j\in I_{00}$
\begin{align}E&\left[\frac{I\left(P_{1j}\leq
\frac{|\mathcal{R}_1(P_1)|c_1(q)q}{m}, j\in \mathcal{R}_1(P_1)
\right)}{\max(|\mathcal{R}_1(P_1)|, 1)}
\right]\notag\\&=\sum_{r=1}^m\frac{1}{r}\textmd{Pr}\left(P_{1j}\leq
\frac{rc_1(q)q}{m}, j\in \mathcal{R}_1(P_1),
|\mathcal{R}_1^{(j)}(P_1^{(j)})|=r-1
\right)\label{foritem3}\\&\leq\sum_{r=1}^m\frac{1}{r}\textmd{Pr}\left(P_{1j}\leq
\frac{rc_1(q)q}{m},  |\mathcal{R}_1^{(j)}(P_1^{(j)})|=r-1
\right)\label{needfor3}\\&\leq
\frac{c_1(q)q}{m}\sum_{r=1}^m\textmd{Pr}\left(|\mathcal{R}_1^{(j)}(P_1^{(j)})|=r-1
\right)=\frac{c_1(q)q}{m}.\label{uncondlast}
\end{align}
The inequality in (\ref{uncondlast}) follows from the independence
of the $p$-values within the primary study and the fact that
$P_{1j}$ is a null-hypothesis $p$-value. The equality in
(\ref{uncondlast}) follows from the fact that $\cup_{r=1}^m
\{|\mathcal{R}_1^{(j)}(P_1^{(j)})|=r-1\}$ is the entire space of
$P_1^{(j)},$ represented as a union of disjoint events. Combining
(\ref{uncond1}) with (\ref{uncondlast}) we obtain for $j\in I_{00}$
\begin{eqnarray}E\left(\frac{R_j}{\max(R,1)}\right)\leq
c_2q\frac{c_1(q)q}{m}.\label{end00prds}\end{eqnarray} Summing this
upper bound over all $j\in I_{00}$ we obtain the upper bound in
(\ref{end00}), thus completing the proof of item 1 for the case
where the set of $p$-values within the follow-up study has property
PRDS.

%The FDR for
%replicability analysis therefore satisfies The FDR for replicability
%analysis therefore satisfies
%\begin{align*}
% FDR&=E\left(\sum_{j\in
%I_{00}}\frac{R_j}{\max(R,1)}\right)+E\left(\sum_{j\in
%I_{01}}\frac{R_j}{\max(R,1)}\right)+E\left(\sum_{j\in
%I_{10}}\frac{R_j}{\max(R,1)}\right) \\&\leq
%f_{00}c_1c_2q^2+f_{01}c_1q+c_2qE\left(\frac{|I_{10}\cap
%\mathcal{R}_1|}{\max(|\mathcal{R}_1|, 1)}\right),
%\end{align*}
%where the inequality follows from summing the upper bound in
%(\ref{end00prds}) over all $j\in I_{00}$ and  using inequalities
%(\ref{fin01}), (\ref{fin10}), thus completing the proof of item 1
%for the case where the $p$-values within the primary study are
%independent and the set of $p$-values within the follow-up study has
%property PRDS.

We shall now prove item 2. Inequalities (\ref{fin01}) and
(\ref{fin10}) follow from the results in the Supplementary Material
of \cite{Bogomolov12}. Specifically, inequality (\ref{fin01})
follows from the derivations leading to (S2.8) and inequality
(\ref{fin10}) follows from the proof of item 2 and item 3 of Lemma
S2.1 in \cite{Bogomolov12}. We shall now prove inequality (\ref{end00}). Both for the case
where the $p$-values within the follow-up study are independent and
for the case where the $p$-values within the follow-up study have
property PRDS, the derivations leading to (\ref{uncond1}) and
(\ref{needfor3}) remain valid when $m$ is replaced with
$m^*$ in those derivations and in the terms defining $C_r^{(j)}.$ %It
%follows that
%\begin{align}
%E\left(\sum_{j\in I_{00}}\frac{R_j}{\max(R,1)}\right)\leq
%c_2q\sum_{j\in I_{00}}E\left[\frac{I\left(P_{1j}\leq
%\frac{|\mathcal{R}_1(P_1)|c_1q}{m^*}, j\in \mathcal{R}_1(P_1)
%\right)}{\max(|\mathcal{R}_1(P_1)|, 1)} \right].\label{uncondmstar}
%\end{align}
%Note that
Therefore
\begin{align}\sum_{j\in I_{00}}E&\left[\frac{I\left(P_{1j}\leq
\frac{|\mathcal{R}_1(P_1)|c_1(q)q}{m^*}, j\in \mathcal{R}_1(P_1)
\right)}{\max(|\mathcal{R}_1(P_1)|, 1)} \right]\notag\\&\leq
\sum_{j\in I_{00}}\sum_{r=1}^m\frac{1}{r}\textmd{Pr}\left(P_{1j}\leq
\frac{rc_1(q)q}{m^*}, |\mathcal{R}_1^{(j)}(P_1^{(j)})|=r-1
\right).\label{cometojasa}
\end{align}
It follows from the derivations leading from (S2.3) to (S2.8) in the
Supplementary Material of \cite{Bogomolov12}, replacing $I_0$ with
$I_{00},$ $q_1$ with $c_1(q)q,$  and the event $C_r^{(j)}$ with the
event $|\mathcal{R}_1^{(j)}(P_1^{(j)})|=r-1$ both in the derivations
and in the definition of $p_{jrl},$  that
\begin{align}
\sum_{j\in I_{00}}\sum_{r=1}^m\frac{1}{r}\textmd{Pr}\left(P_{1j}\leq
\frac{rc_1(q)q}{m^*},
|\mathcal{R}_1^{(j)}(P_1^{(j)})|=r-1\right)\leq
|I_{00}|\frac{c_1(q)q}{m}.\label{fromjasa}
\end{align}
Combining (\ref{uncond1}) with $m$ replaced by $m^*,$
(\ref{cometojasa}) and (\ref{fromjasa}) we obtain inequality
(\ref{end00}), which completes the proof of item 2.

We shall now prove item 3. If we replace  $\widetilde{q}_1$ with
$\widetilde{c}_1(q)q$ and $|I_0|$ with $|I_{01}|$ in the derivations
leading to (S2.18) in the Supplementary Material in
\cite{Bogomolov12}, we obtain
\begin{align}E\left(\sum_{j\in I_{01}}\frac{R_j}{\max(R,1)}\right)\leq|I_{01}|\frac{\widetilde{c}_1(q)q}{m}=f_{01}\widetilde{c}_1(q)q.\label{fin01new}\end{align}
It follows from the definition of $\widetilde{c}_1(x)$ that
$\widetilde{c}_1(x)\leq c_1(x)$ for all $x\in(0,1),$ in particular
$\widetilde{c}_1(q)\leq c_1(q).$ Inequality (\ref{fin01}) follows
immediately from this inequality and inequality (\ref{fin01new}).
Inequality (\ref{fin10}) is obtained using the derivations from the
main manuscript and the Supplementary Material of
\cite{Bogomolov12}, as detailed in the proof of inequality
(\ref{fin10}) in item 1. For this item $q_1$ is replaced with
$\widetilde{c}_1(q)q$ in those derivations,  and in order to obtain
inequality (\ref{fin10}) we use the fact that
$\widetilde{c}_1(q)\leq c_1(q).$ We shall now prove inequality
(\ref{end00}). Both for the case where the $p$-values within the
follow-up study are independent and for the case where the
$p$-values within the follow-up study have property PRDS, the
derivations leading to (\ref{uncond1}) and (\ref{foritem3}) remain
valid when $c_1(q)$ is replaced with $\widetilde{c}_1(q)$ in those
derivations and in the terms defining $C_r^{(j)}.$ Therefore
\begin{align}\sum_{j\in I_{00}}E\left(\frac{R_j}{\max(R,1)}\right)\leq c_2q\sum_{j\in I_{00}}\sum_{r=1}^m\frac{1}{r}\textmd{Pr}\left(P_{1j}\leq
\frac{r\widetilde{c}_1(q)q}{m}, j\in \mathcal{R}_1(P_1),
|\mathcal{R}_1^{(j)}(P_1^{(j)})|=r-1 \right).\label{fin31}
\end{align}
It follows from the derivations leading from (S2.9) to (S2.18) in
the Supplementary Material of \cite{Bogomolov12}, replacing $I_0$
with $I_{00},$ $\widetilde{q}_1$ with $\widetilde{c}_1(q)q,$  and
the event $C_r^{(j)}$ with the event
$|\mathcal{R}_1^{(j)}(P_1^{(j)})|=r-1$ both in the derivations and
in the definition of $\widetilde{p}_{jrl},$  that
\begin{align}
\sum_{j\in I_{00}}\sum_{r=1}^m\frac{1}{r}\textmd{Pr}\left(P_{1j}\leq
\frac{r\widetilde{c}_1(q)q}{m}, j\in \mathcal{R}_1(P_1),
|\mathcal{R}_1^{(j)}(P_1^{(j)})|=r-1 \right)\leq
|I_{00}|\frac{\widetilde{c}_1(q)q}{m}.\label{fin32}
\end{align}
Combining (\ref{fin31}) with (\ref{fin32}), and using the fact that
$\widetilde{c}_1(q)\leq c_1(q),$ we obtain inequality (\ref{end00}),
which completes the proof of item 3.

\begin{lemma}\label{lemmathm1}
For Steps 1-4 in the computation of $r$-values: %and
%$\widetilde{c}_1(x)$ as defined in item 3 of Theorem 1:
\begin{enumerate}
\item For feature $i\in \mathcal{R}_1,$ if there exists a solution $r_i\in(0,1)$ to $f_i(r_i)=r_i,$ then this solution is
unique, i.e. the $r$-value in Step 4 is well-defined.
\item Item 1 holds when the function $f_i(x)$ is computed with the modification
in item 2 of Theorem 1.
\item Declaring the features with $r$-values at most $q$ is
equivalent to the procedure given in Section Variations (left column
at the bottom of page 4).%\footnote{In the proof of Theorem 1, we
%consider the procedure given in the left column at the bottom of
%page 4, which will be always for brevity referred to as "procedure
%in Section Variations".}
%\item %For feature $i\in \mathcal{R}_1,$ the solution to $\widetilde{f}_i(r_i)=r_i$ is
%unique, where $\widetilde{f}_i(x)$ is obtained by replacing $c_1(x)$
%by $\widetilde{c}_1(x)$ in Step 3.
\item For $r$-values computed with the modification in item 2 of Theorem 1, declaring the features with $r$-values at most $q$ is
equivalent to the procedure given in Section Variations where $m$
 is replaced by $m^*=m\sum_{i=1}^m1/i.$
\item The function $\widetilde{c}_1(x)$ in item 3 of Theorem 1 is well-defined.
For $r$-values computed with the modification in item 3 of Theorem 1, declaring the features with
$r$-values at most $q$ is equivalent to the procedure given in
Section Variations where $c_1(q)$ is replaced by $\tilde{c}_1(q).$
%\item When $c_1(x)$ is replaced by $\widetilde{c}_1(x)$ in Step 3, declaring the features with
%$r$-values at most $q$ is equivalent to the procedure given in
%Section Variations where $c_1(q)$ is replaced by $\tilde{c}_1(q).$
\end{enumerate}
\end{lemma}
\textbf{Proof of Lemma \ref{lemmathm1}}.
\\\\\textbf{Proof of items 1 and 2 of Lemma \ref{lemmathm1}.} Simple calculations show that
$g(x)=xc_1(x)$ is a strictly increasing function of $x$ for $x>0.$
Therefore for each feature $j\in \mathcal{R}_1,$ $e_j(x)/x$ is a
strictly decreasing function of $x.$ Despite the fact that
$e_j(x)/[x\cdot rank(e_j(x))]$ may not be monotone decreasing
functions for $j\in \mathcal{R}_1,$ it is guaranteed that
$f_i(x)/x=\min_{\{j: e_j(x)\geq e_i(x), j\in
\mathcal{R}_1\}}e_j(x)/[x\cdot rank(e_j(x))]$ is a strictly
decreasing function of $x$ for each feature $i\in
\mathcal{R}_1.$\footnote{The proof that $f_i(x)/x$ is a strictly
decreasing function is quite involved and is omitted for brevity.}
Therefore if there exists a solution $r_i\in(0,1)$ to $f_i(x)/x=1,$
then it is unique, since for all $x<r_i,$ $f_i(x)/x>1$ and for all
$x>r_i,$ $f_i(x)/x<1.$ When the function $f_i(x)$ is computed with
the modification in item 2 of Theorem 1, the proof remains the same,
since $m^*$ does not depend on $x.$
\\\\\textbf{Proof of items 3-5 of Lemma \ref{lemmathm1}.} It is easy to see that for the
procedure given in Section Variations, $\mathcal{R}_2=\{i\in
\mathcal{R}_1: f_i(q)\leq q\}.$ The same result holds for the
function $f_i(x)$ with the modification of item 2 and item 3 of
Theorem 1 and the modified procedures in items 4 and 5 of Lemma
\ref{lemmathm1}, respectively. Therefore it is enough to prove that
for $i\in \mathcal{R}_1,$ $f_i(q)\leq q$ if and only if $r_i\leq q$
for each one of the items of Lemma \ref{lemmathm1}.
\\\textbf{Proof of item 3.}
Assume $f_i(q)\leq q.$ Note that $f_i(x)$ can be defined on $[0,1)$
and $f_i(0)>0$ since the $p$-values are positive. It can be shown
that $f_i(x)$ is a continuous function on $[0,1),$\footnote{It is
easy to see that $f_i(x)$ is continuous at each $x_0$ where
$e$-values are unique. Note that for each $j\in \mathcal{R}_1$  the
numerator of $e_j(x)m/rank(e_j(x))$ is continuous and there is a
small neighbourhood  of $x_0$ where $rank(e_j(x))$ does not change,
yielding that $e_j(x)m/rank(e_j(x))$ is continuous at $x_0.$ Since
the minimum of continuous functions is also continuous, $f_i(x)$ is
a continuous function as well. For $x_0$ where $e$-values are not
unique, the proof is more involved. In these points the functions
$e_j(x)m/rank(e_j(x))$ may be not continuous, however $f_i(x)$ is
continuous. }
%when there is at least one positive $p$-value. Note that $f_i(x)$ is,
therefore $h_i(x)=f_i(x)-x$ is a continuous function as well. Using
the facts that $h_i(0)=f_i(0)-0>0$ and $h_i(q)=f_i(q)- q\leq 0,$ we
obtain from the intermediate value theorem that there exists a value
$0<x_i\leq q$ satisfying $f_i(x_i)=x_i.$ Using item 1 we obtain that
this solution is unique and $r_i=x_i.$ Thus we have proved $r_i\leq
q.$ Let us now assume that $r_i\leq q$ and prove that $f_i(q)\leq
q.$ Since $r_i\leq q,$ $r_i\neq 1,$ therefore $r_i$ is the unique
solution in $(0, 1)$ to $f_i(x)=x.$ It follows from the fact that
$f_i(x)/x$ is monotone decreasing (see proof of item 1) that
$f_i(q)/q\leq f_i(r_i)/r_i=1,$ therefore $f_i(q)\leq q.$
\\\textbf{Proof of item
4.} We need to prove that when we replace $m$ with
$m^*=m\sum_{i=1}^m1/i$ in the computation of $f_i(x)$ and $r_i$ for
$i\in \mathcal{R}_1,$ $r_i\leq q$ if and only if feature $i$ is
rejected by the procedure in Section Variations, where $m$ is
replaced by $m=m^*.$ It is easy to see that for this modified
procedure, $\mathcal{R}_2=\{i\in \mathcal{R}_1: f_i(q)\leq q\},$
where $f_i(q)$ is computed with the modification above. It remains
to prove that $\{i\in \mathcal{R}_1:f_i(q)\leq q\}=\{i \in
\mathcal{R}_1: r_i\leq q\}.$ Since $f_i(x)$ is continuous, it is
obvious that the modified function $f_i(x)$ is continuous as well.
Moreover, $f_i(x)/x$ is monotone decreasing in $x$, thus using
arguments similar to the proof of item 3 the result follows.
\\\textbf{Proof of item
5.} The proof that the function $\widetilde{c}_1(x)$ is
well-defined, i.e. that for all $x\in(0,1)$ there exists a solution
$a$ to $a\sum_{i=1}^{\lceil tm/(ax)-1 \rceil}1/i=c_1(x)$ is
technical and therefore is omitted. Similarly to the items above, we
need to prove that $\{i\in \mathcal{R}_1:\widetilde{f}_i(q)\leq
q\}=\{i \in \mathcal{R}_1: \widetilde{r}_i\leq q\},$ where
$\widetilde{f}_i(x)$ and $\widetilde{r}_i$ are the modified
functions and $r$-values respectively, given in item 3 of Theorem 1.
We shall first show that if $\widetilde{f}_i(q)\leq q,$ then there
exists $\widetilde{r}_i=\min\{x:\widetilde{f}_i(x)\leq x\}\in(0,1).$
It can be shown that $\widetilde{c}_1(x)$ is right
continuous,\footnote{The proof that $\widetilde{c}_1(x)$ is right
continuous is based on the facts that $\lceil tm/(ax)-1 \rceil$ is a
right continuous function of $x$ and $c_1(x)$ is a continuous
function. Since the proof is technical, it is omitted.} and
therefore $\tilde{f}_i(x)$ is right continuous. If
$\widetilde{f}_i(q)\leq q,$ then $\inf\{x:\tilde{f}_i(x)\leq x\}<1.$
It remains to show that $\inf\{x:\tilde{f}_i(x)\leq x\}\neq 0,$
since $\tilde{f}_i(x)$ is right continuous for all $x\in(0,1)$,
therefore if $\inf\{x:\tilde{f}_i(x)\leq x\}\in(0,1),$ then
$\inf\{x:\tilde{f}_i(x)\leq x\}=\min\{x:\tilde{f}_i(x)\leq x\}.$ We
shall now prove that $\inf\{x:\tilde{f}_i(x)\leq x\}\neq 0.$ Note
that $\widetilde{c}_1(x)\leq c_1(x)$ for all $x\in(0,1),$ therefore
it can be shown that $\tilde{f}_i(x)\geq f_i(x)$ for all
$x\in(0,1).$ As we noticed in the proof of item 3 of Lemma
\ref{lemmathm1}, $f_i(x)$ can be defined for $x\in[0,1)$, it is a
continuous function on $[0,1)$, and $f_i(0)>0.$ Therefore there
exists $\delta>0$ such that $f_i(x)>x$ for $x\in[0,\delta).$ It
follows that $\tilde{f}_i(x)>x$ for $x\in(0,\delta),$ therefore
$\inf\{x:\tilde{f}_i(x)\leq x\}\neq 0.$ Thus we have proved that if
$\widetilde{f}_i(q)\leq q,$ then there exists
$\widetilde{r}_i=\min\{x:\widetilde{f}_i(x)\leq x\}\in(0,1).$ From
the definition of $\widetilde{r}_i$ we obtain $\widetilde{r}_i\leq
q.$ Assume now that $\widetilde{r}_i\leq q,$ i.e.
$\min\{x:\widetilde{f}_i(x)\leq x\}\leq q.$ It can be shown that
$\widetilde{f}_i(x)/x$ is a monotone decreasing
function,\footnote{The proof that $\widetilde{f}_i(x)/x$ is a
monotone decreasing function is quite involved and is omitted for
brevity. It is based on the facts that $\widetilde{c}_1(x)x$ is
monotone increasing, therefore $e_j(x)/x$ is monotone decreasing for
all $j\in \mathcal{R}_1.$ } therefore
$\widetilde{f}_i(q)/q\leq\widetilde{f}_i(r_i)/r_i\leq 1,$ i.e.
$\widetilde{f}_i(q)\leq q.$

%%%%%%%%%%%%%%%%%%%%%%%%%%%%%%%%%%%%%%%%%%%%%%%%%%%%%%%%%%%%%%%%

%% Adding Figure and Table References
%% Be sure to add figures and tables after \end{article}
%% and before \end{document}

%% For figures, put the caption below the illustration.
%%
%% \begin{figure}
%% \caption{Almost Sharp Front}\label{afoto}
%% \end{figure}

%% For Tables, put caption above table
%%
%% Table caption should start with a capital letter, continue with lower case
%% and not have a period at the end
%% Using @{\vrule height ?? depth ?? width0pt} in the tabular preamble will
%% keep that much space between every line in the table.

%% \begin{table}
%% \caption{Repeat length of longer allele by age of onset class}
%% \begin{tabular}{@{\vrule height 10.5pt depth4pt  width0pt}lrcccc}
%% table text
%% \end{tabular}
%% \end{table}

%% For two column figures and tables, use the following:

%% \begin{figure*}
%% \caption{Almost Sharp Front}\label{afoto}
%% \end{figure*}

%% \begin{table*}
%% \caption{Repeat length of longer allele by age of onset class}
%% \begin{tabular}{ccc}
%% table text
%% \end{tabular}
%% \end{table*}

\section{GWAS Real data examples}

Tables \ref{tab-ex1}, \ref{tab-ex3}, and \ref{tab-ex4} show the results of the
replicability analysis for the SNPs followed-up based on the results
of the primary study (or studies). Columns 1-3 in Tables \ref{tab-ex1} and \ref{tab-ex3} and columns  1-2 in Table \ref{tab-ex4} contain the  position
of each SNP. Columns 4-5 in Tables \ref{tab-ex1} and \ref{tab-ex3} and columns  3-4 in Table \ref{tab-ex4} show the primary and follow-up $p$-values.
Columns  6-8 in Tables \ref{tab-ex1} and \ref{tab-ex3} and column 6 in Table \ref{tab-ex4}show the $r$-values for different choices of
$l_{00}$. Column 9 in Tables \ref{tab-ex1} and \ref{tab-ex3} and column 5 in Table \ref{tab-ex4} shows the meta-analysis $p$-values, which
are the unadjusted $p$-values computed using the data from the
primary and follow-up studies for testing the global null hypothesis
of no association in any of the studies. In Tables \ref{tab-ex1} and \ref{tab-ex3} the rows are sorted by the
meta-analysis $p$-values, and the handful of findings with most
significant meta-analysis $p$-values which were reported as
interesting in the published works are marked with an $*$ in the
last column.
% latex table generated in R 2.15.0 by xtable 1.7-0 package
% Tue Nov 20 14:36:20 2012
\newpage

\begin{center}
\scriptsize
\begin{longtable}{p{0.2cm}p{1.4cm}p{1.6cm}p{1.6cm}p{1.6cm}p{1.2cm}p{1.2cm}p{1.2cm}p{1.2cm}p{0.1cm}}
\caption{Replicability analysis for the study of \cite{Yu12}%, with $444882$ SNPs in the primary study, and 61 SNPs followed-up
.}\label{tab-ex1}
\endfirsthead
  \hline
  Chr.  & Position & Gene & p1 & p2 & $l_{00}=0$ & $l_{00}=0.5$& $l_{00}=0.8$&  p\_meta &  \\
  \hline
  6 & 32685358 & HLA-DRB1 & 8.19e-08 & 8.57e-14 & 0.0243 & 0.0150 & 0.0074 & 4.13e-20 & * \\
  8 & 6810195 & DEFAs & 2.04e-07 & 1.25e-07 & 0.0409 & 0.0207 & 0.0090 & 3.18e-14 & * \\
  6 & 32779226 & HLA-DQA/B & 3.28e-08 & 3.57e-06 & 0.0224 & 0.0147 & 0.0059 & 3.43e-13 & * \\
  22 & 28753460 & MTMR3 & 2.30e-07 & 2.02e-05 & 0.0409 & 0.0207 & 0.0090 & 1.17e-11 & * \\
  6 & 30049922 & HLA-A & 4.05e-09 & 3.68e-04 & 0.0224 & 0.0150 & 0.0090 & 1.74e-11 & * \\
  17 & 7403693 & TNFSF13 & 1.50e-06 & 2.52e-05 & 0.1907 & 0.1001 & 0.0413 & 9.40e-11 & * \\
  17 & 7431901 & MPDU1 & 5.52e-07 & 3.16e-04 & 0.0819 & 0.0418 & 0.0169 & 4.31e-10 & * \\
  2 & 111315937 & ACOXL & 6.83e-05 & 3.41e-03 &   1 &   1 &   1 & 4.08e-07 &  \\
  16 & 31255249 & x & 6.67e-05 & 7.41e-03 &   1 &   1 &   1 & 4.64e-06 &  \\
  4 & 78121177 & x & 3.14e-10 & 8.16e-01 &   1 &   1 &   1 & 2.23e-05 &  \\
  11 & 113369319 & x & 1.82e-09 & 9.74e-01 &   1 &   1 &   1 & 5.42e-05 &  \\
  7 & 33386800 & BBS9 & 2.75e-05 & 1.67e-01 &   1 &   1 &   1 & 1.17e-04 &  \\
  11 & 44042263 & x & 1.74e-05 & 2.72e-01 &   1 &   1 &   1 & 1.24e-04 &  \\
  4 & 40144579 & x & 9.95e-07 & 6.72e-01 &   1 &   1 &   1 & 1.85e-04 &  \\
  12 & 13229380 & x & 1.23e-05 & 4.41e-01 &   1 &   1 &   1 & 3.09e-04 &  \\
  14 & 69116920 & x & 4.60e-05 & 3.72e-01 &   1 &   1 &   1 & 3.71e-04 &  \\
  8 & 30305114 & x & 3.19e-05 & 4.73e-01 &   1 &   1 &   1 & 5.38e-04 &  \\
  12 & 129587780 & x & 4.59e-05 & 5.53e-01 &   1 &   1 &   1 & 6.84e-04 &  \\
  6 & 31382359 & x & 8.20e-08 & 9.53e-01 &   1 &   1 &   1 & 7.64e-04 &  \\
  16 & 77632003 & WWOX & 7.20e-05 & 4.57e-01 &   1 &   1 &   1 & 1.04e-03 &  \\
  8 & 97393458 & PTDSS1 & 5.67e-05 & 6.12e-01 &   1 &   1 &   1 & 1.09e-03 &  \\
  6 & 26384629 & x & 4.32e-06 & 2.79e-01 &   1 &   1 &   1 & 1.25e-03 &  \\
  13 & 62434248 & x & 3.77e-05 & 5.39e-01 &   1 &   1 &   1 & 1.70e-03 &  \\
  11 & 109836841 & FDX1 & 7.15e-05 & 7.19e-01 &   1 &   1 &   1 & 2.03e-03 &  \\
  18 & 35923102 & x & 4.35e-05 & 2.85e-01 &   1 &   1 &   1 & 2.32e-03 &  \\
  6 & 13733392 & RANBP9 & 1.70e-05 & 8.45e-01 &   1 &   1 &   1 & 2.55e-03 &  \\
  9 & 78162069 & PSAT1 & 5.98e-05 & 8.01e-01 &   1 &   1 &   1 & 2.71e-03 &  \\
  10 & 55006847 & x & 7.93e-05 & 6.87e-01 &   1 &   1 &   1 & 3.16e-03 &  \\
  6 & 33163516 & x & 1.46e-04 & 7.74e-01 &   1 &   1 &   1 & 4.56e-03 &  \\
  7 & 158006056 & x & 9.26e-05 & 7.50e-01 &   1 &   1 &   1 & 4.99e-03 &  \\
  6 & 106231017 & x & 6.19e-05 & 8.59e-01 &   1 &   1 &   1 & 6.41e-03 &  \\
  21 & 19339830 & x & 7.81e-05 & 5.34e-01 &   1 &   1 &   1 & 6.58e-03 &  \\
  12 & 19488937 & AEBP2 & 4.95e-05 & 4.77e-01 &   1 &   1 &   1 & 6.92e-03 &  \\
  18 & 57221085 & x & 8.62e-06 & 5.48e-01 &   1 &   1 &   1 & 7.96e-03 &  \\
  10 & 76538473 & DUSP13 & 7.84e-05 & 8.54e-01 &   1 &   1 &   1 & 9.18e-03 &  \\
  8 & 1307131 & x & 4.95e-05 & 7.52e-01 &   1 &   1 &   1 & 1.14e-02 &  \\
  16 & 72315398 & x & 4.92e-05 & 9.92e-01 &   1 &   1 &   1 & 1.24e-02 &  \\
  3 & 130747968 & H1FOO & 1.85e-05 & 8.90e-01 &   1 &   1 &   1 & 1.65e-02 &  \\
  12 & 39245441 & x & 1.21e-07 & 4.38e-01 &   1 &   1 &   1 & 1.66e-02 &  \\
  7 & 92588411 & CCDC132 & 1.28e-07 & 4.09e-01 &   1 &   1 &   1 & 1.77e-02 &  \\
  1 & 110389963 & x & 1.46e-07 & 2.59e-01 &   1 &   1 &   1 & 1.99e-02 &  \\
  9 & 21342862 & x & 7.95e-05 & 9.45e-01 &   1 &   1 &   1 & 2.18e-02 &  \\
  2 & 46170592 & PRKCE & 1.78e-05 & 3.40e-01 &   1 &   1 &   1 & 2.21e-02 &  \\
  17 & 52636364 & x & 3.45e-05 & 5.20e-01 &   1 &   1 &   1 & 2.26e-02 &  \\
  1 & 82547439 & x & 5.51e-05 & 8.89e-01 &   1 &   1 &   1 & 2.72e-02 &  \\
  6 & 156238397 & x & 4.73e-05 & 1.43e-01 &   1 &   1 &   1 & 2.80e-02 &  \\
  11 & 61956393 & x & 2.16e-06 & 5.37e-01 &   1 &   1 &   1 & 4.10e-02 &  \\
  10 & 135319919 & x & 3.90e-05 & 6.76e-01 &   1 &   1 &   1 & 4.33e-02 &  \\
  12 & 66026196 & x & 2.57e-06 & 5.08e-01 &   1 &   1 &   1 & 4.42e-02 &  \\
  8 & 25535212 & x & 2.46e-05 & 3.44e-01 &   1 &   1 &   1 & 5.31e-02 &  \\
  15 & 88817746 & IQGAP1 & 8.64e-05 & 2.22e-01 &   1 &   1 &   1 & 5.76e-02 &  \\
  6 & 13707282 & SIRT5 & 3.98e-05 & 3.66e-01 &   1 &   1 &   1 & 6.84e-02 &  \\
  1 & 70907559 & x & 3.96e-05 & 4.71e-01 &   1 &   1 &   1 & 7.24e-02 &  \\
  1 & 176696794 & CEP350 & 7.14e-05 & 4.50e-01 &   1 &   1 &   1 & 1.04e-01 &  \\
  12 & 8955888 & x & 7.85e-06 & 2.07e-01 &   1 &   1 &   1 & 1.10e-01 &  \\
  11 & 94090071 & x & 5.22e-05 & 3.08e-01 &   1 &   1 &   1 & 1.29e-01 &  \\
  2 & 4641380 & x & 9.57e-05 & 3.68e-01 &   1 &   1 &   1 & 1.39e-01 &  \\
  1 & 23749819 & x & 8.10e-05 & 2.08e-01 &   1 &   1 &   1 & 1.58e-01 &  \\
  7 & 105466371 & x & 4.61e-05 & 9.90e-02 &   1 &   1 &   1 & 2.32e-01 &  \\
  5 & 4489013 & x & 8.96e-05 & 3.83e-02 &   1 &   1 &   1 & 4.40e-01 &  \\
  1 & 215993345 & x & 2.67e-05 & 1.32e-02 &   1 &   1 &   1 & 4.90e-01 &  \\
   \hline
\end{longtable}
\end{center}

%\newpage

\begin{center}
\scriptsize
\begin{longtable}{p{0.2cm}p{0.2cm}p{1.4cm}p{1.6cm}p{1.2cm}p{1.2cm}p{1.2cm}p{1.2cm}p{1.2cm}p{0.1cm}}
\caption{Replicability analysis for the study of \cite{Barrett08},
with 635547 SNPs in the primary study, and with 126 SNPs
followed-up. The last column marks the 30 SNPs that were highlighted
as ``convincingly (Bonferroni $P <0.05$) replicated CD risk loci",
based on the follow-up study $p$-values, in Table 2 of the main
manuscript of \cite{Barrett08}. }\label{tab-ex3}
\endfirsthead
  \hline
row \# & Chr.  & Position & p1 & p2 & $l_{00}=0$ & $l_{00}=0.5$& $l_{00}=0.8$ &  p\_meta &  \\
  \hline
1 & 1 & 67417979 & 3.19e-34 & 1.50e-36 & 4.05e-28 & 2.03e-28 & 8.11e-29 & 2.15e-68 &  \\
  2 & 1 & 67414547 & 5.05e-36 & 3.10e-29 & 3.91e-27 & 3.91e-27 & 3.91e-27 & 3.33e-63 & * \\
  3 & 1 & 67387537 & 1.35e-24 & 5.62e-17 & 4.72e-15 & 4.72e-15 & 4.72e-15 & 1.82e-39 &  \\
  4 & 2 & 233962410 & 5.66e-21 & 7.67e-14 & 4.83e-12 & 4.83e-12 & 4.83e-12 & 1.18e-32 & * \\
  5 & 5 & 40428485 & 2.51e-22 & 2.79e-08 & 1.34e-06 & 1.14e-06 & 1.14e-06 & 3.09e-27 & * \\
  6 & 5 & 40437266 & 2.26e-22 & 3.18e-08 & 1.34e-06 & 1.14e-06 & 1.14e-06 & 3.41e-27 &  \\
  7 & 2 & 233965368 & 1.28e-21 & 3.66e-05 & 5.76e-04 & 5.76e-04 & 5.76e-04 & 4.61e-25 &  \\
  8 & 10 & 64108492 & 9.51e-12 & 1.61e-10 & 1.73e-06 & 1.14e-06 & 4.84e-07 & 2.23e-20 & * \\
  9 & 5 & 131798704 & 2.29e-09 & 3.52e-11 & 2.08e-04 & 1.04e-04 & 4.16e-05 & 1.16e-18 & * \\
  10 & 18 & 12769947 & 5.95e-12 & 2.41e-07 & 7.20e-06 & 6.48e-06 & 6.48e-06 & 2.55e-17 & * \\
  11 & 10 & 101281583 & 8.53e-11 & 1.69e-07 & 1.05e-05 & 6.48e-06 & 5.32e-06 & 1.53e-16 & * \\
  12 & 5 & 150239060 & 3.18e-11 & 2.57e-07 & 7.20e-06 & 6.48e-06 & 6.48e-06 & 1.70e-16 & * \\
  13 & 10 & 101282445 & 9.09e-11 & 3.10e-07 & 1.05e-05 & 7.10e-06 & 7.10e-06 & 3.05e-16 &  \\
  14 & 18 & 12799340 & 3.27e-11 & 1.23e-06 & 2.38e-05 & 2.38e-05 & 2.38e-05 & 7.05e-16 &  \\
  15 & 5 & 150203580 & 4.09e-11 & 7.47e-07 & 1.57e-05 & 1.57e-05 & 1.57e-05 & 7.33e-16 &  \\
  16 & 13 & 43355925 & 8.04e-08 & 1.33e-07 & 5.68e-03 & 2.84e-03 & 1.21e-03 & 1.04e-13 & * \\
  17 & 5 & 158747111 & 4.40e-09 & 3.66e-06 & 3.73e-04 & 1.86e-04 & 7.46e-05 & 1.93e-13 & * \\
  18 & 6 & 167408399 & 1.65e-07 & 3.26e-07 & 8.74e-03 & 4.57e-03 & 1.91e-03 & 5.22e-13 & * \\
  19 & 3 & 49696536 & 1.08e-07 & 5.64e-07 & 6.54e-03 & 3.27e-03 & 1.38e-03 & 5.76e-13 & * \\
  20 & 17 & 37767727 & 2.97e-06 & 9.15e-08 & 8.58e-02 & 4.60e-02 & 2.07e-02 & 3.41e-12 & * \\
  21 & 3 & 49676987 & 9.47e-08 & 2.24e-06 & 6.02e-03 & 3.01e-03 & 1.27e-03 & 3.55e-12 &  \\
  22 & 1 & 197667523 & 3.41e-07 & 2.34e-06 & 1.44e-02 & 7.50e-03 & 3.36e-03 & 7.17e-12 & * \\
  23 & 12 & 39104262 & 8.95e-08 & 6.55e-05 & 5.99e-03 & 3.00e-03 & 1.27e-03 & 6.36e-11 &  \\
  24 & 6 & 106541962 & 1.85e-06 & 7.70e-06 & 6.03e-02 & 3.23e-02 & 1.42e-02 & 1.22e-10 & * \\
  25 & 9 & 114645994 & 1.96e-07 & 6.58e-05 & 9.46e-03 & 4.93e-03 & 2.18e-03 & 1.30e-10 & * \\
  26 & 12 & 38888207 & 6.64e-08 & 1.65e-04 & 4.96e-03 & 2.49e-03 & 1.91e-03 & 1.54e-10 & * \\
  27 & 6 & 20836710 & 1.26e-07 & 2.78e-04 & 7.28e-03 & 3.65e-03 & 2.92e-03 & 4.48e-10 & * \\
  28 & 11 & 75978964 & 7.16e-08 & 7.32e-04 & 8.02e-03 & 6.83e-03 & 6.36e-03 & 6.60e-10 & * \\
  29 & 21 & 44439989 & 5.41e-06 & 1.59e-05 & 1.32e-01 & 7.27e-02 & 3.18e-02 & 7.04e-10 & * \\
  30 & 1 & 157665119 & 1.75e-07 & 4.81e-04 & 8.90e-03 & 4.93e-03 & 4.33e-03 & 7.30e-10 & * \\
  31 & 1 & 169593891 & 2.01e-07 & 3.21e-04 & 9.46e-03 & 4.93e-03 & 3.24e-03 & 7.66e-10 & * \\
  32 & 1 & 197691964 & 9.69e-07 & 1.00e-04 & 3.52e-02 & 1.94e-02 & 8.07e-03 & 8.10e-10 &  \\
  33 & 10 & 35327656 & 4.24e-06 & 2.53e-05 & 1.10e-01 & 6.03e-02 & 2.64e-02 & 8.93e-10 & * \\
  34 & 19 & 1074378 & 5.80e-09 & 3.47e-03 & 2.82e-02 & 2.57e-02 & 2.19e-02 & 1.06e-09 &  \\
  35 & 19 & 1075031 & 6.48e-09 & 2.10e-02 & 1.10e-01 & 9.80e-02 & 8.97e-02 & 1.18e-09 &  \\
  36 & 20 & 61798026 & 7.60e-07 & 1.38e-04 & 2.93e-02 & 1.57e-02 & 6.52e-03 & 1.30e-09 &  \\
  37 & 7 & 50081722 & 1.58e-05 & 9.41e-06 & 2.73e-01 & 1.67e-01 & 8.20e-02 & 1.39e-09 &  \\
  38 & 6 & 167405736 & 1.65e-07 & 1.21e-03 & 1.09e-02 & 1.02e-02 & 9.24e-03 & 1.58e-09 &  \\
  39 & 9 & 4971602 & 3.40e-07 & 4.30e-04 & 1.44e-02 & 7.50e-03 & 4.01e-03 & 1.73e-09 & * \\
  40 & 6 & 32789255 & 1.53e-08 & 3.82e-03 & 2.93e-02 & 2.75e-02 & 2.35e-02 & 2.17e-09 &  \\
  41 & 8 & 126609233 & 2.45e-06 & 1.09e-04 & 7.41e-02 & 3.87e-02 & 1.74e-02 & 2.25e-09 & * \\
  42 & 7 & 50046933 & 2.46e-05 & 1.10e-05 & 3.68e-01 & 2.36e-01 & 1.24e-01 & 2.30e-09 & * \\
  43 & 17 & 35294289 & 1.06e-06 & 2.92e-04 & 3.74e-02 & 2.06e-02 & 8.57e-03 & 2.50e-09 & * \\
  44 & 6 & 32484449 & 7.23e-09 & 6.02e-03 & 4.10e-02 & 3.79e-02 & 3.23e-02 & 2.60e-09 &  \\
  45 & 8 & 126603853 & 1.90e-06 & 1.82e-04 & 6.04e-02 & 3.23e-02 & 1.42e-02 & 2.78e-09 &  \\
  46 & 9 & 114648320 & 1.31e-07 & 4.22e-03 & 3.13e-02 & 2.95e-02 & 2.53e-02 & 3.67e-09 &  \\
  47 & 21 & 15727091 & 1.03e-05 & 4.58e-05 & 2.02e-01 & 1.16e-01 & 5.37e-02 & 3.70e-09 & * \\
  48 & 1 & 114015850 & 7.75e-06 & 8.25e-05 & 1.67e-01 & 9.75e-02 & 4.28e-02 & 4.95e-09 &  \\
  49 & 1 & 114089610 & 9.05e-06 & 1.01e-04 & 1.89e-01 & 1.10e-01 & 4.85e-02 & 7.30e-09 & * \\
  50 & 10 & 35589263 & 6.05e-06 & 1.76e-04 & 1.40e-01 & 8.00e-02 & 3.42e-02 & 8.04e-09 &  \\
  51 & 21 & 44436378 & 5.21e-06 & 3.61e-04 & 1.30e-01 & 7.14e-02 & 3.16e-02 & 1.43e-08 &  \\
  52 & 21 & 15734423 & 1.00e-05 & 4.44e-04 & 2.02e-01 & 1.16e-01 & 5.31e-02 & 3.36e-08 &  \\
  53 & 3 & 49499240 & 2.42e-08 & 1.94e-01 & 5.28e-01 & 5.28e-01 & 5.28e-01 & 3.56e-08 &  \\
  54 & 9 & 4978761 & 1.96e-06 & 1.62e-03 & 6.08e-02 & 3.25e-02 & 1.42e-02 & 4.34e-08 &  \\
  55 & 2 & 61129193 & 3.07e-06 & 2.80e-03 & 8.67e-02 & 4.64e-02 & 2.08e-02 & 6.36e-08 &  \\
  56 & 1 & 169594596 & 1.90e-07 & 2.60e-02 & 1.30e-01 & 1.16e-01 & 1.09e-01 & 9.01e-08 &  \\
  57 & 3 & 49425868 & 2.84e-08 & 1.07e-01 & 3.28e-01 & 3.16e-01 & 3.16e-01 & 1.20e-07 &  \\
  58 & 13 & 43497789 & 6.90e-07 & 8.82e-03 & 5.85e-02 & 5.05e-02 & 4.36e-02 & 1.44e-07 &  \\
  59 & 2 & 61098480 & 3.82e-06 & 5.65e-03 & 1.03e-01 & 5.54e-02 & 3.16e-02 & 1.57e-07 &  \\
  60 & 6 & 20797924 & 1.83e-07 & 2.88e-02 & 1.34e-01 & 1.19e-01 & 1.17e-01 & 1.64e-07 &  \\
  61 & 6 & 5096246 & 3.54e-07 & 1.92e-02 & 1.03e-01 & 9.49e-02 & 8.34e-02 & 3.48e-07 &  \\
  62 & 1 & 157691986 & 2.98e-07 & 2.77e-02 & 1.32e-01 & 1.16e-01 & 1.14e-01 & 3.82e-07 &  \\
  63 & 17 & 29611838 & 2.01e-06 & 1.35e-02 & 7.91e-02 & 7.14e-02 & 6.19e-02 & 5.34e-07 &  \\
  64 & 17 & 37824128 & 7.42e-06 & 7.40e-03 & 1.65e-01 & 9.50e-02 & 4.17e-02 & 7.10e-07 &  \\
  65 & 19 & 18300383 & 5.43e-08 & 5.26e-02 & 2.02e-01 & 1.92e-01 & 1.84e-01 & 7.54e-07 &  \\
  66 & 2 & 27652888 & 3.62e-05 & 3.81e-03 & 5.06e-01 & 3.16e-01 & 1.75e-01 & 1.15e-06 &  \\
  67 & 6 & 3378317 & 1.04e-06 & 3.91e-02 & 1.67e-01 & 1.54e-01 & 1.49e-01 & 1.37e-06 &  \\
  68 & 2 & 102521887 & 1.02e-05 & 1.60e-02 & 2.02e-01 & 1.16e-01 & 7.20e-02 & 1.45e-06 &  \\
  69 & 2 & 27642591 & 3.44e-05 & 1.08e-02 & 4.86e-01 & 3.14e-01 & 1.70e-01 & 2.30e-06 &  \\
  70 & 2 & 230934834 & 7.59e-06 & 5.44e-02 & 2.02e-01 & 1.93e-01 & 1.85e-01 & 2.48e-06 &  \\
  71 & 20 & 61820069 & 2.04e-07 & 3.30e-01 & 7.58e-01 & 7.58e-01 & 7.58e-01 & 2.66e-06 &  \\
  72 & 6 & 3379241 & 1.15e-06 & 5.82e-02 & 2.13e-01 & 2.04e-01 & 1.96e-01 & 2.83e-06 &  \\
  73 & 10 & 75302766 & 1.23e-05 & 3.14e-02 & 2.23e-01 & 1.32e-01 & 1.24e-01 & 3.03e-06 &  \\
  74 & 1 & 7840274 & 1.47e-06 & 5.41e-02 & 2.02e-01 & 1.93e-01 & 1.85e-01 & 3.63e-06 &  \\
  75 & 6 & 149618772 & 3.64e-06 & 4.40e-02 & 1.85e-01 & 1.67e-01 & 1.64e-01 & 4.39e-06 &  \\
  76 & 6 & 21578398 & 4.97e-06 & 6.78e-02 & 2.41e-01 & 2.34e-01 & 2.25e-01 & 5.02e-06 &  \\
  77 & 22 & 20264229 & 1.25e-06 & 3.25e-01 & 7.58e-01 & 7.58e-01 & 7.58e-01 & 6.26e-06 &  \\
  78 & 11 & 63906946 & 4.74e-06 & 2.45e-01 & 6.30e-01 & 6.30e-01 & 6.30e-01 & 7.44e-06 &  \\
  79 & 4 & 187576360 & 1.35e-06 & 8.65e-02 & 2.87e-01 & 2.83e-01 & 2.76e-01 & 7.81e-06 &  \\
  80 & 2 & 230916728 & 8.93e-06 & 8.43e-02 & 2.83e-01 & 2.80e-01 & 2.72e-01 & 9.04e-06 &  \\
  81 & 17 & 29849794 & 1.25e-05 & 9.61e-02 & 3.10e-01 & 3.07e-01 & 2.99e-01 & 1.01e-05 &  \\
  82 & 2 & 102529086 & 1.08e-05 & 4.93e-02 & 2.02e-01 & 1.83e-01 & 1.75e-01 & 1.11e-05 &  \\
  83 & 20 & 57351084 & 1.73e-06 & 1.01e-01 & 3.22e-01 & 3.14e-01 & 3.10e-01 & 1.18e-05 &  \\
  84 & 4 & 187585769 & 1.34e-06 & 1.07e-01 & 3.28e-01 & 3.16e-01 & 3.16e-01 & 1.33e-05 &  \\
  85 & 16 & 84545499 & 4.74e-06 & 2.26e-01 & 5.87e-01 & 5.87e-01 & 5.87e-01 & 1.40e-05 &  \\
  86 & 18 & 17927329 & 1.59e-05 & 4.43e-02 & 2.73e-01 & 1.67e-01 & 1.64e-01 & 1.44e-05 &  \\
  87 & 18 & 54054001 & 5.56e-06 & 2.07e-01 & 5.55e-01 & 5.55e-01 & 5.55e-01 & 1.97e-05 &  \\
  88 & 14 & 75071147 & 4.71e-06 & 1.52e-01 & 4.35e-01 & 4.26e-01 & 4.26e-01 & 2.25e-05 &  \\
  89 & 5 & 37949301 & 1.74e-06 & 2.73e-01 & 6.68e-01 & 6.68e-01 & 6.68e-01 & 2.41e-05 &  \\
  90 & 10 & 75324937 & 1.12e-05 & 1.04e-01 & 3.28e-01 & 3.16e-01 & 3.16e-01 & 3.32e-05 &  \\
  91 & 6 & 21565929 & 1.09e-05 & 1.23e-01 & 3.68e-01 & 3.56e-01 & 3.56e-01 & 3.40e-05 &  \\
  92 & 11 & 63967228 & 1.60e-05 & 8.82e-02 & 2.89e-01 & 2.85e-01 & 2.78e-01 & 3.45e-05 &  \\
  93 & 12 & 58059725 & 2.84e-05 & 1.49e-01 & 4.32e-01 & 4.22e-01 & 4.22e-01 & 3.97e-05 &  \\
  94 & 22 & 20281207 & 8.65e-07 & 4.93e-01 & 1.00e+00 & 1.00e+00 & 1.00e+00 & 4.55e-05 &  \\
  95 & 4 & 106463957 & 6.25e-06 & 2.71e-01 & 6.68e-01 & 6.68e-01 & 6.68e-01 & 5.03e-05 &  \\
  96 & 1 & 222692358 & 2.73e-06 & 3.93e-01 & 8.46e-01 & 8.46e-01 & 8.46e-01 & 5.08e-05 &  \\
  97 & 4 & 7649390 & 3.24e-06 & 3.52e-01 & 7.99e-01 & 7.99e-01 & 7.99e-01 & 5.27e-05 &  \\
  98 & 17 & 35315722 & 3.41e-06 & 4.19e-01 & 8.95e-01 & 8.95e-01 & 8.95e-01 & 5.45e-05 &  \\
  99 & 3 & 133674827 & 6.84e-06 & 1.61e-01 & 4.56e-01 & 4.46e-01 & 4.46e-01 & 6.21e-05 &  \\
  100 & 1 & 7766478 & 1.45e-05 & 2.11e-01 & 5.60e-01 & 5.60e-01 & 5.60e-01 & 6.82e-05 &  \\
  101 & 8 & 83235127 & 1.32e-05 & 2.23e-01 & 5.85e-01 & 5.85e-01 & 5.85e-01 & 1.28e-04 &  \\
  102 & 21 & 39215894 & 8.73e-06 & 2.63e-01 & 6.63e-01 & 6.63e-01 & 6.63e-01 & 1.65e-04 &  \\
  103 & 10 & 122495603 & 2.08e-05 & 3.63e-01 & 8.10e-01 & 8.10e-01 & 8.10e-01 & 1.98e-04 &  \\
  104 & 14 & 75056332 & 1.28e-05 & 3.31e-01 & 7.58e-01 & 7.58e-01 & 7.58e-01 & 2.22e-04 &  \\
  105 & 13 & 80961793 & 1.61e-07 & 3.72e-01 & 8.22e-01 & 8.22e-01 & 8.22e-01 & 2.23e-04 &  \\
  106 & 18 & 75866208 & 1.38e-06 & 2.56e-01 & 6.52e-01 & 6.52e-01 & 6.52e-01 & 2.80e-04 &  \\
  107 & 12 & 13070503 & 8.89e-06 & 4.35e-01 & 9.18e-01 & 9.18e-01 & 9.18e-01 & 3.27e-04 &  \\
  108 & 10 & 132842492 & 2.65e-05 & 4.41e-01 & 9.18e-01 & 9.18e-01 & 9.18e-01 & 3.88e-04 &  \\
  109 & 5 & 37948752 & 1.06e-05 & 4.41e-01 & 9.18e-01 & 9.18e-01 & 9.18e-01 & 4.78e-04 &  \\
  110 & 13 & 80973593 & 2.64e-07 & 3.29e-02 & 1.48e-01 & 1.32e-01 & 1.28e-01 & 5.54e-04 &  \\
  111 & 12 & 13046606 & 4.05e-05 & 4.55e-01 & 9.40e-01 & 9.40e-01 & 9.40e-01 & 7.51e-04 &  \\
  112 & 10 & 1453158 & 7.72e-06 & 2.90e-01 & 6.96e-01 & 6.96e-01 & 6.96e-01 & 7.56e-04 &  \\
  113 & 1 & 181883035 & 1.04e-05 & 3.88e-01 & 8.43e-01 & 8.43e-01 & 8.43e-01 & 9.32e-04 &  \\
  114 & 18 & 59311578 & 1.01e-05 & 4.98e-01 & 1.00e+00 & 1.00e+00 & 1.00e+00 & 9.48e-04 &  \\
  115 & 7 & 130385443 & 1.24e-05 & 3.81e-01 & 8.35e-01 & 8.35e-01 & 8.35e-01 & 9.64e-04 &  \\
  116 & 18 & 55030807 & 1.75e-05 & 3.04e-01 & 7.23e-01 & 7.23e-01 & 7.23e-01 & 1.06e-03 &  \\
  117 & 19 & 50999246 & 1.95e-05 & 4.60e-01 & 9.42e-01 & 9.42e-01 & 9.42e-01 & 1.32e-03 &  \\
  118 & 15 & 72660732 & 7.44e-06 & 2.73e-01 & 6.68e-01 & 6.68e-01 & 6.68e-01 & 1.33e-03 &  \\
  119 & 18 & 55028896 & 8.34e-06 & 3.56e-01 & 8.01e-01 & 8.01e-01 & 8.01e-01 & 1.80e-03 &  \\
  120 & 16 & 84542932 & 3.44e-04 & 2.78e-01 & 1.00e+00 & 1.00e+00 & 1.00e+00 & 2.39e-03 &  \\
  121 & 8 & 107779719 & 2.92e-05 & 3.14e-01 & 7.40e-01 & 7.40e-01 & 7.40e-01 & 2.78e-03 &  \\
  122 & 12 & 58052436 & 1.14e-05 & 2.89e-01 & 6.96e-01 & 6.96e-01 & 6.96e-01 & 3.45e-03 &  \\
  123 & 18 & 54054701 & 9.40e-06 & 1.95e-01 & 5.28e-01 & 5.28e-01 & 5.28e-01 & 5.28e-03 &  \\
  124 & 18 & 75865061 & 2.11e-06 & 1.08e-01 & 3.28e-01 & 3.16e-01 & 3.16e-01 & 6.92e-03 &  \\
  125 & 15 & 72685472 & 5.81e-06 & 7.71e-02 & 2.70e-01 & 2.59e-01 & 2.52e-01 & 1.27e-02 &  \\
  126 & 8 & 107743073 & 2.59e-05 & 1.43e-01 & 4.19e-01 & 4.09e-01 & 4.09e-01 & 1.36e-02 &  \\
   \hline
\end{longtable}
\end{center}

\begin{table}[!hbp]
\begin{center}
\caption{Replicability analysis for FWER control for the study of \cite{Cheung12} on GWAS of TPP.
The number of SNPs in the primary study was 486782, and four SNPs
were followed-up. The lower bound for $f_{00}$ was $l_{00}=0.8$ for the $r$-value computation. }\label{tab-ex4}
\begin{tabular}{llllll}
  \hline
  Chr. & Position & p1 & p2 & p\_meta & $r$-value\\
  \hline
 17 & 65837933 & 6.28e-10 & 1.49e-05 & 7.69e-14  & 0.00012\\
 17 & 65818432 & 1.39e-09 & 7.36e-05 & 1.59e-12  & 0.00059 \\
 17 & 65799923 & 2.27e-09 & 7.25e-05 & 1.09e-12  & 0.00058 \\
 17 & 65778654 & 1.84e-08 & 0.000116 & 1.6e-11  & 0.00360\\
   \hline
\end{tabular}
\end{center}
\end{table}

\section{Choice of selection rule for replicability analysis} Although any stable selection rule can be used, some selection rules may be more efficient than others. For a given
FDR level $q$, the promising hypotheses for
replicability analysis are the set of hypotheses
rejected with the BH procedure at level $c_1(q)q$ on the primary study
$p$-values. Therefore, for the purpose of replicability analysis,
the set of hypotheses to be considered should be only this set or
a subset thereof. This means that if $R_1$ hypotheses are
followed-up, not all $R_1$ features need to be
selected for a replicability analysis at a predetermined level $q$.
The advantage of selecting only the relevant subset is that the power of the procedure will be greater since the problem of multiplicity among the selected will be smaller, without compromising any potential replicability claims. Specifically, in order for the $r$-value to be below $q$, only the
subset of $R_1$ hypotheses selected for follow-up with primary study
$p$-values that are small enough need to be considered, where our
requirement for small enough is as follows: when applying the BH procedure at level $c_1(q)q$ on $p_{11}, \ldots, p_{1m}$, these hypotheses will be among the rejected.
%that the  BH adjusted $p$-values of
%the $R_1$ primary study $p$-values, corrected for multiplicity of
%$m$ hypotheses, are below $c_1q$. %Formally, if SNP $i\in
%\mathcal{R}_1$, it should be considered for replicability analysis
%only if $\min_{p_{1j}\geq p_{1i}, j\in \mathcal{R}_1}
%\frac{mp_{1j}}{rank(p_{1j})}\leq c_1q$, where $rank(p_{1j})$ is the rank of $p_{1j}$ among all primary study $p$-values (with maximum rank for ties).
 Computing the $r$-values for the subset of $\mathcal{R}_1$
with small enough primary study $p$-values, we receive smaller
$r$-values than if all $R_1$ SNPs are considered for replicability
analysis.

For the example of GWAS of IgA nephropathy, for an FDR level of 0.05, only 14
SNPs out of the  61 followed-up had primary study $p$-values small
enough to be considered for replicability analysis.
 The number of $r$-values below
0.05 was still seven with this modified selection rule, but these
seven $r$-values were smaller than the $r$-values for the seven SNPs
in Table 1 of the main manuscript. Specifically, with parameters $(l_{00},c_2)
= (0.8, 0.5)$ for this superior selection rule that selected 14 SNPs
for follow-up, the $r$-values were 0.005, 0.008, 0.005, 0.008,
0.005, 0.041, 0.017, whereas the $r$-values computed using  all 61
SNPs selected were, respectively, 0.007, 0.009, 0.006, 0.009, 0.009,
0.041, and 0.017.
%For example 3, for an FDR level of 0.05, only 117 SNPs out of
%the the 129 SNPs followed-up should be considered for replicability
%analysis. The number of $r$-values below 0.05 remained 54 with this
%modified selection rule.

\section{Power comparison for different values of $(l_{00},c_2)$}\label{app-simulations}
%TO DO: DESCRIPTION OF THE FIGURES.
We conducted simulations in order to investigate how the power and
FDR of our proposal depends on
$c_2\in(0, 1)$ and $l_{00}\in\{0, 0.5, 0.8, 0.9\}$ for $q=0.05.$ %In
%all simulations, in Step 1 of Procedure 1 we used the approximation
%$c_1\approx \frac{1-c_2}{1-l_{00}}$. %so there was no need to prefix
%$q$ for the $r$-value computation.
The $p$-values were generated
independently as follows. Let $P_{1j}$ and $P_{2j}$
be the $p$-values in the primary and in the follow-up
study, respectively, for feature $j$. We set $P_{1j}=1-\Phi(X_{1j})$ and
$P_{2j}=1-\Phi(X_{2j}),$ where $X_{1j}\sim N(\mu_{1j}, 1),$
$X_{2j}\sim N(\mu_{2j}, 1).$ For $i\in\{1,2\},$ we set $\mu_{ij}=0$
if feature $j$ comes from a true null hypothesis in study $i$, and
$\mu_{ij}=\mu_i>0$ if feature $j$ comes from a false null hypothesis in study $i$.
The values of $\mu_1$ and $\mu_2$ were set according to the
requirement that the power of the Bonferroni procedure at level 0.05 in
the primary study is $\pi_1,$ and in the follow-up study is $\pi_2,$
for $\pi_1=0.1$ and $\pi_2\in\{0.2, 0.5, 0.8\}.$ Specifically, we
set $\mu_1=\Phi^{-1}(1-0.05/m)-\Phi^{-1}(1-\pi_1),$ and
$\mu_2=\Phi^{-1}(1-0.05/R_1)-\Phi^{-1}(1-\pi_2),$ where $\Phi^{-1}$
is the inverse of the cumulative distribution function of a standard
normal variable and $R_1$ is the number of rejected hypotheses by
the BH procedure at level $c_1\times 0.05$ applied on the primary
study $p$-values. In addition, $m=1000$, $f_{00}=0.9,$ $f_{01}=f_{10}=0.025,$ $f_{11}=0.05.$

The simulation results were based on 10000 repetitions. The FDR was
estimated by averaging the false discovery proportion. The average
power was estimated by the average number of true replicability
claims, divided by $mf_{11}.$ We also estimated the probability that
our proposal makes at least one true replicability claim
 (which we refer to
as "power for at least one") by the proportion of repetitions in
which at least one true replicability claim was made. The standard
errors of the estimators were of the order
of $10^{-3}$ or $10^{-4}$  for all the sets of parameters. %while the standard error of
%the estimator of power for at least one was %of the order of
%%$10^{-2}$ for some sets of parameters, and was
%always below 0.016.

\begin{figure}[!hpb]\
\centering \subfloat[Average power, $(\pi_1, \pi_2)=(0.1, 0.2).$
]{\label{fig20025:3a}\includegraphics[width=0.5\textwidth,
height=5.5cm]{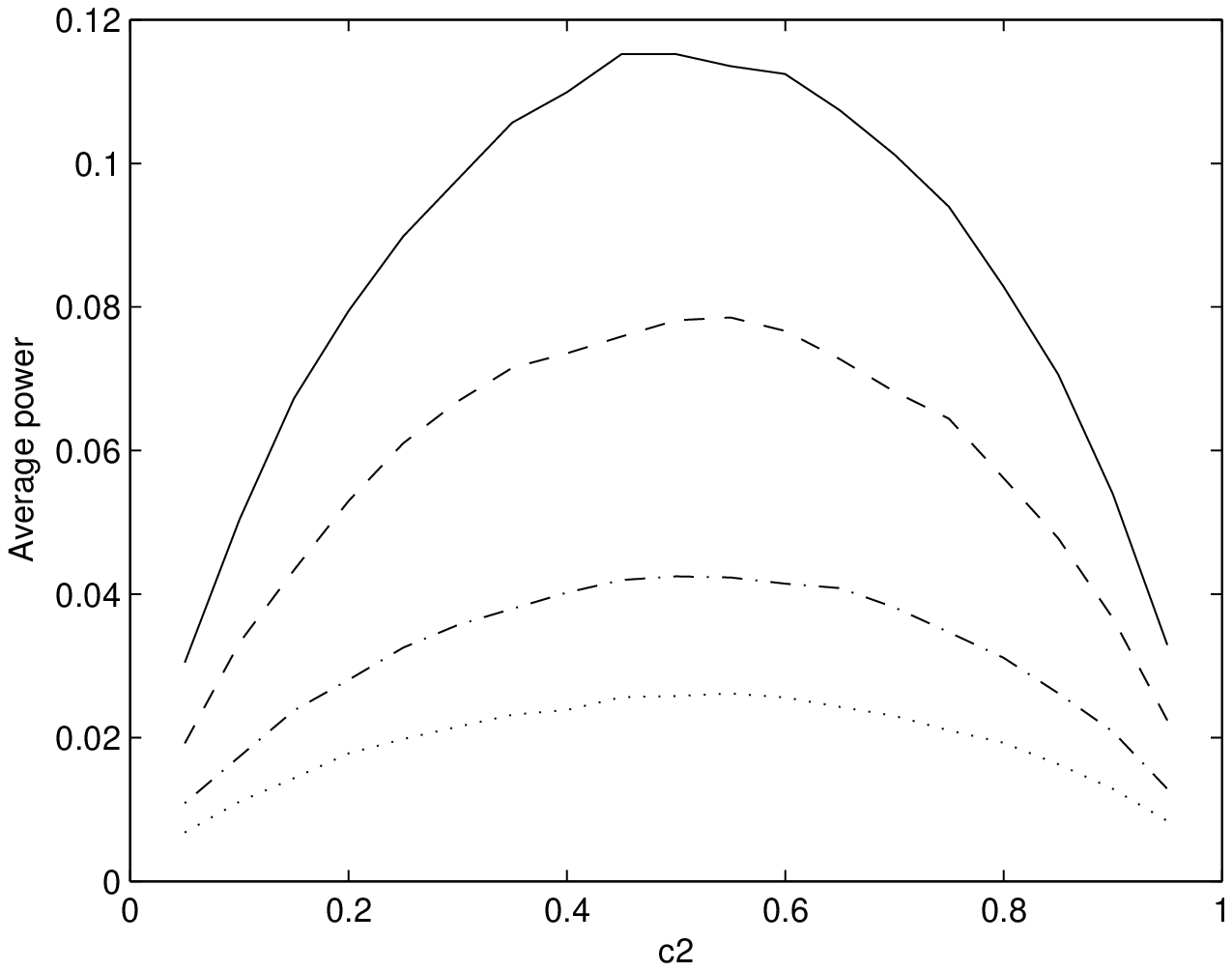}} \subfloat[Power for at least
one, $(\pi_1, \pi_2)=(0.1,
0.2).$]{\label{fig20025:1b}\includegraphics[width=0.5\textwidth,
height=5.5cm]{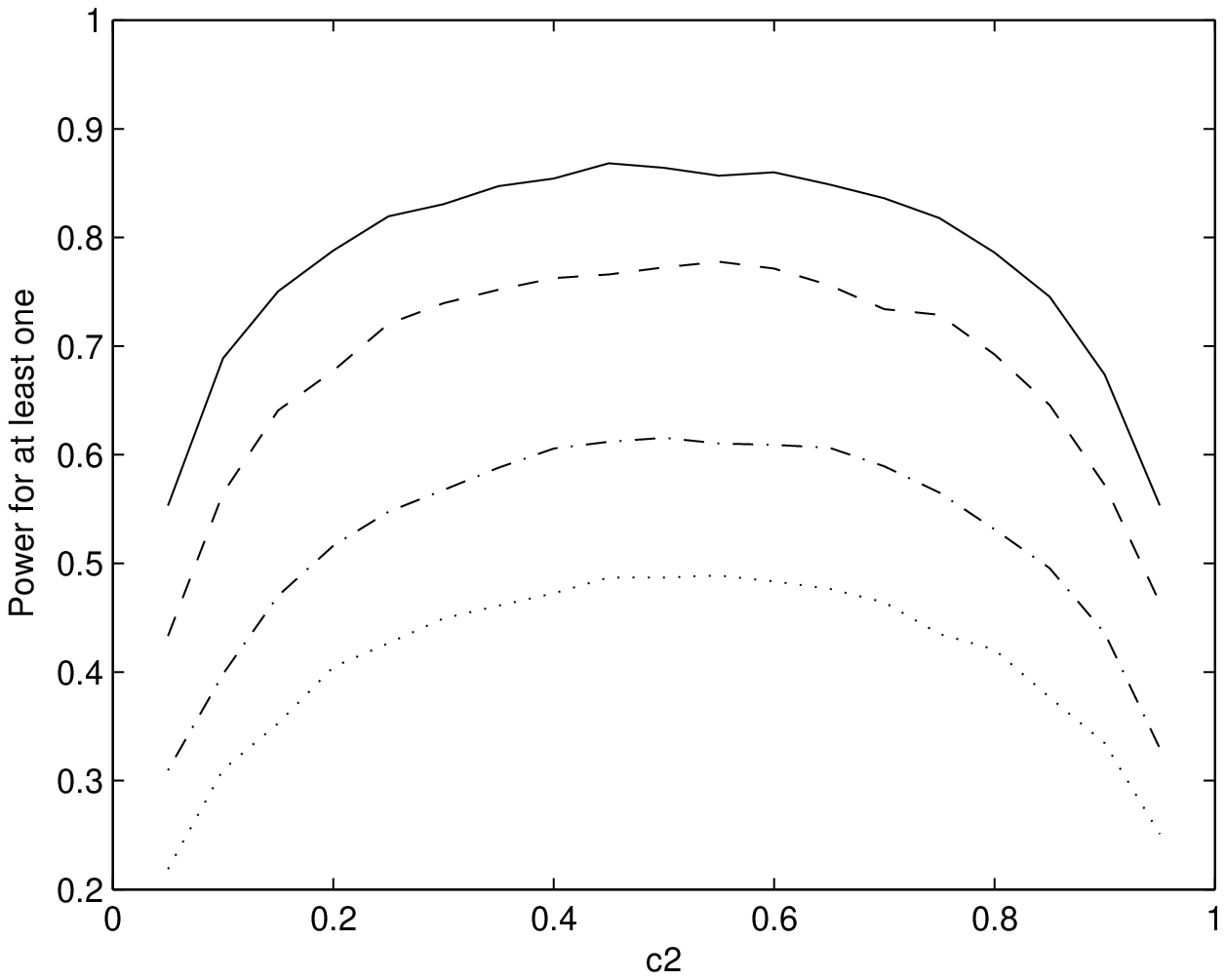}}\\
\subfloat[Average power, $(\pi_1, \pi_2)=(0.1,
0.5).$]{\label{fig20025:3c}\includegraphics[width=0.5\textwidth,
height=5.5cm]{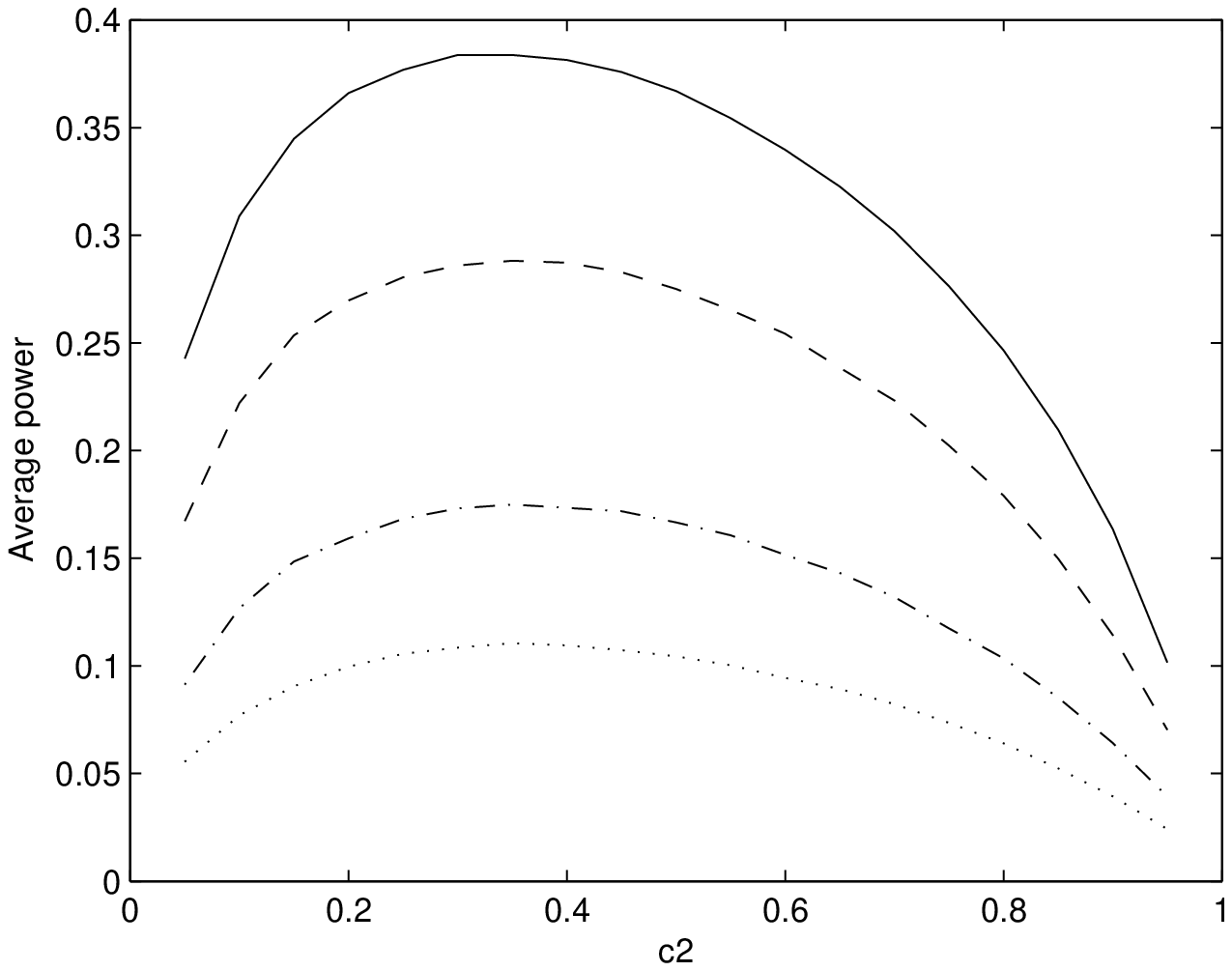}} \subfloat[Power for at least
one, $(\pi_1, \pi_2)=(0.1,
0.5).$]{\label{fig20025:1d}\includegraphics[width=0.5\textwidth,
height=5.5cm]{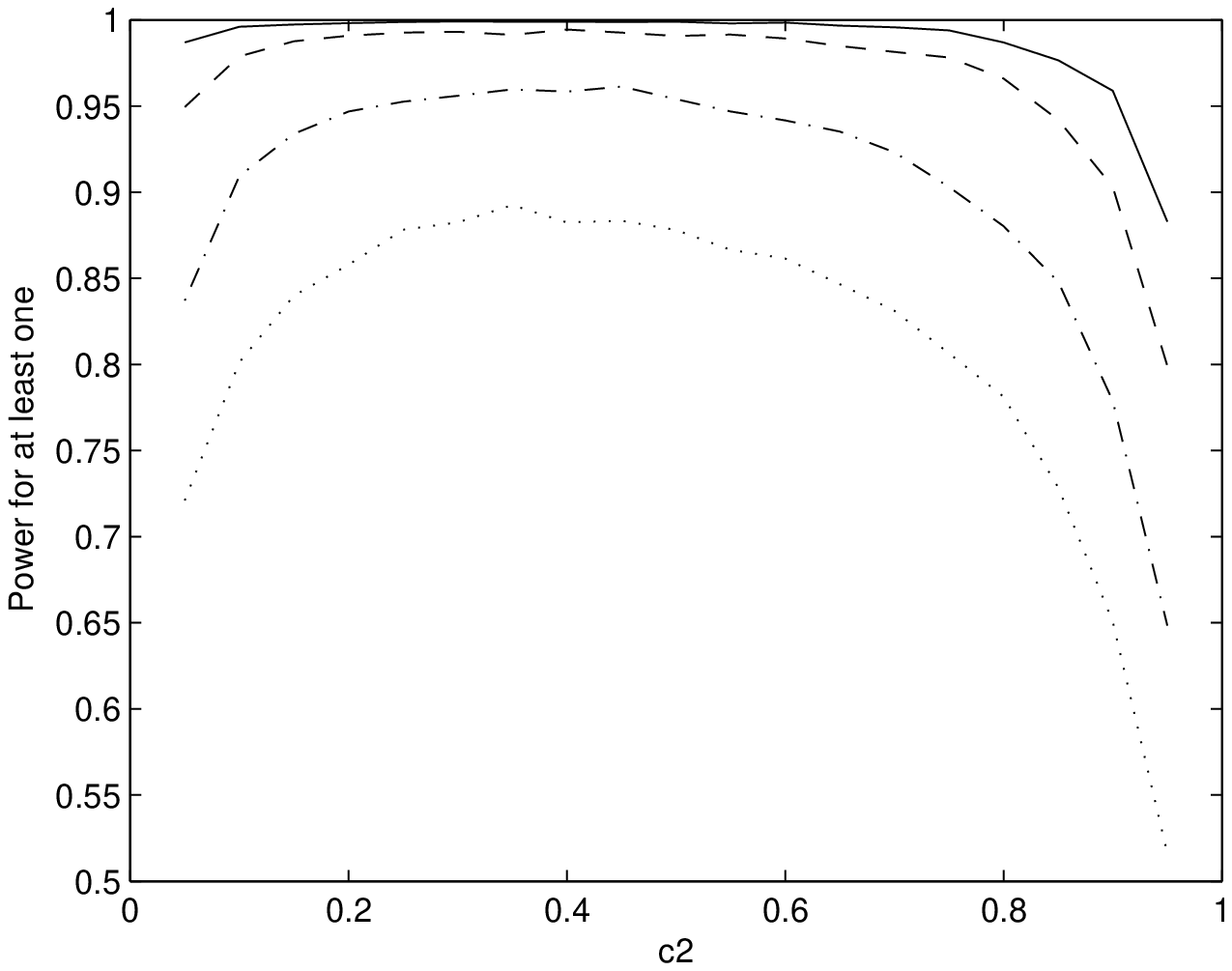}}\\
%\subfloat[FDR]{\label{fig20025:1}\includegraphics[width=0.33\textwidth,
%height=5.5cm]{fdr0102th.eps}}
 \subfloat[Average
power, $(\pi_1, \pi_2)=(0.1,
0.8).$]{\label{fig20025:3e}\includegraphics[width=0.5\textwidth,
height=5.5cm]{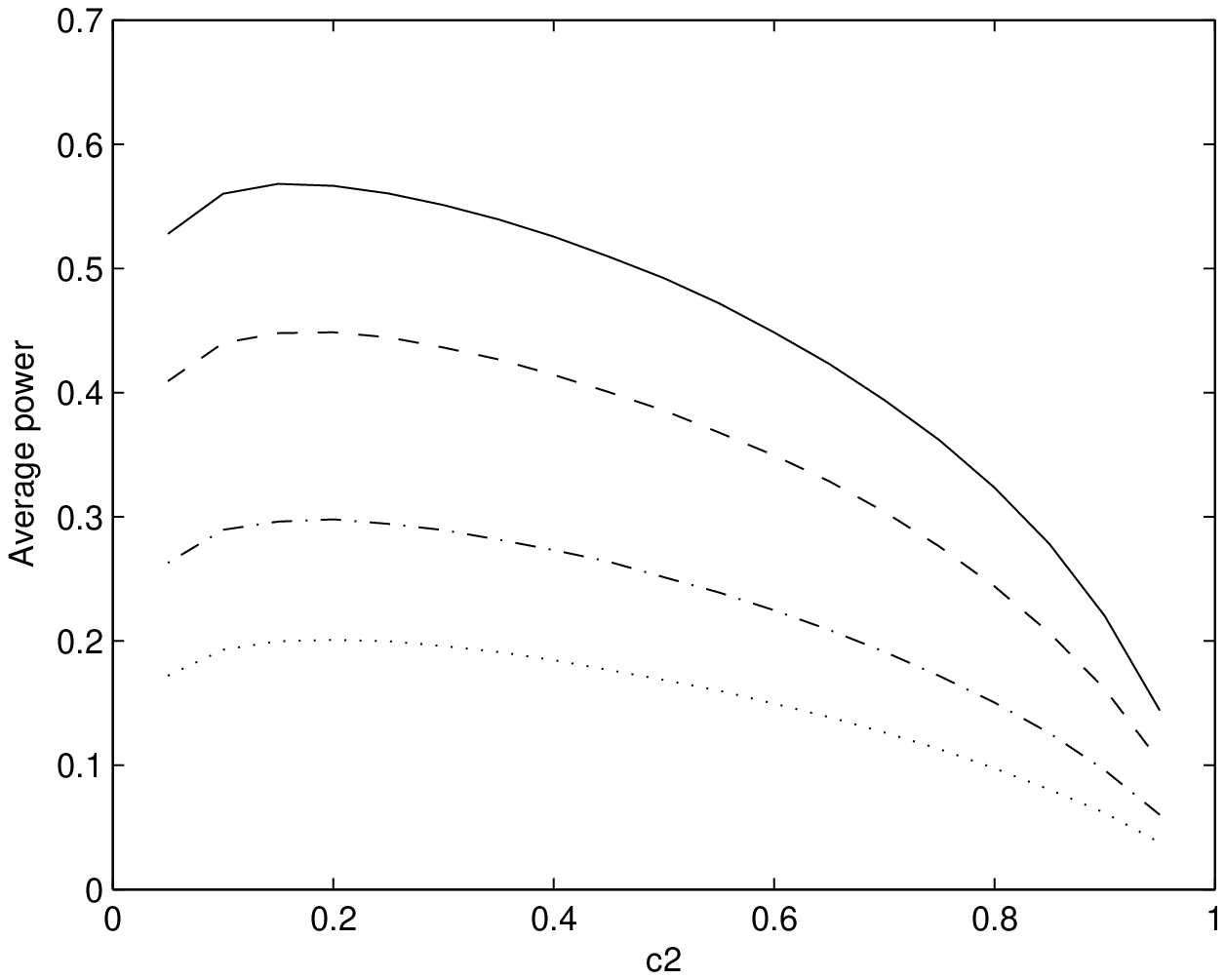}} \subfloat[Power for at least
one, $(\pi_1, \pi_2)=(0.1,
0.8).$]{\label{fig20025:1f}\includegraphics[width=0.5\textwidth,
height=5.5cm]{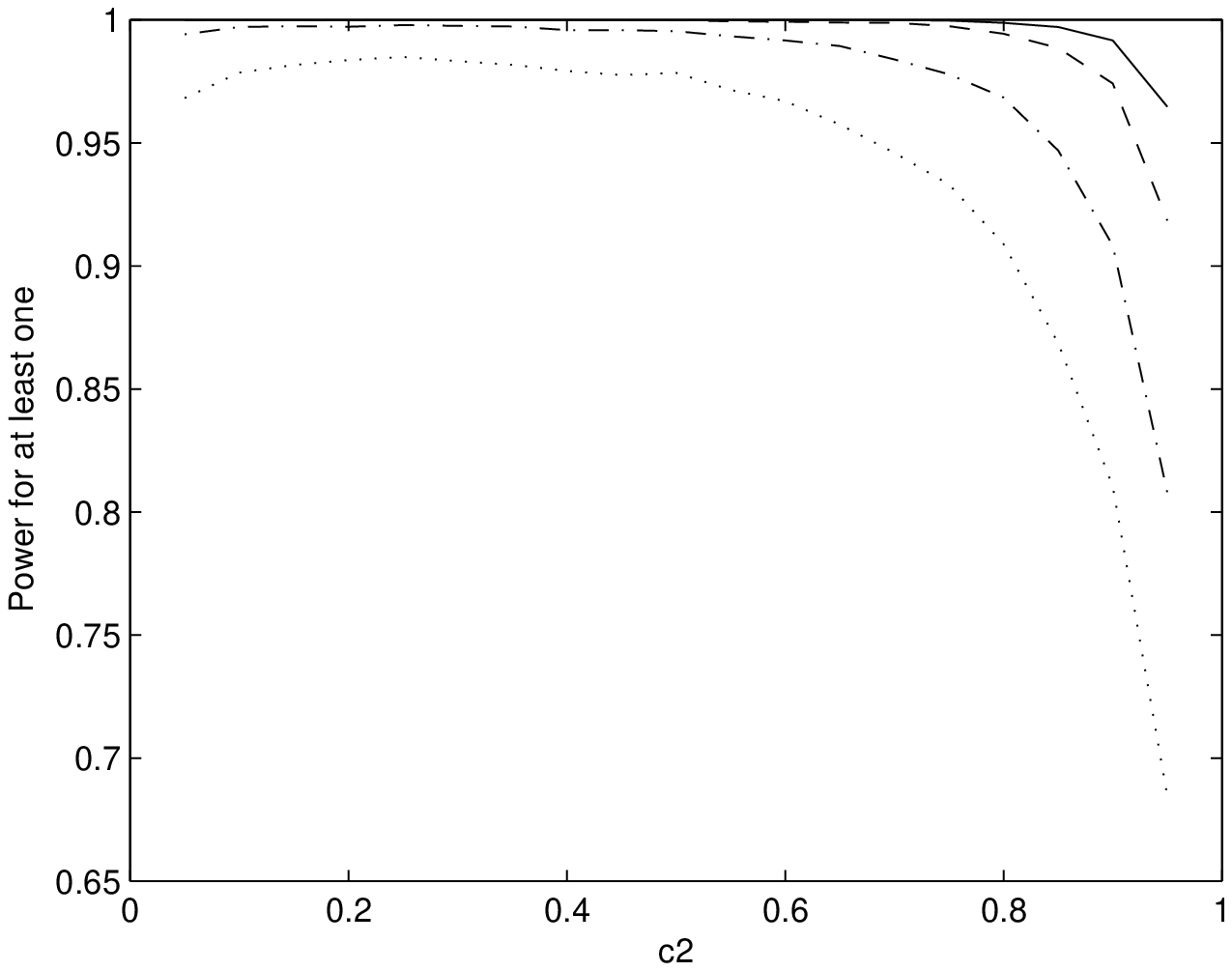}} \caption{The estimated
average power (first column) and the probability of at least one
true replicability claim (power for at least one, column 2) of our
proposal with parameters $(l_{00}, c_2,0.05)$ as a function of $c_2$
in a simulation where $f_{00}=0.9, f_{01}=f_{10}=0.025,
f_{11}=0.05,$ the number of hypotheses examined in the primary study
is 1000, and the signal to noise ratios for the primary study and
the follow-up study,
respectively, are taken according to the requirement that the power
of the Bonferroni procedure at level 0.05 in the primary study is
$\pi_1,$ and in the follow-up study is $\pi_2$ for $(\pi_1,
\pi_2)=(0.1, 0.2)$ (row 1), $(\pi_1, \pi_2)=(0.1, 0.5)$ (row 2),
$(\pi_1, \pi_2)=(0.1, 0.8)$ (row 3); $l_{00}=0.9$ (solid),
$l_{00}=0.8$ (dashed), $l_{00}=0.5$ (dash-dotted), and $l_{00}=0$
(dotted). The standard errors of the estimators were of the order of
$10^{-3}$ or $10^{-4}$ for all the sets of
parameters.}\label{figpower}
\end{figure}

A comparison of columns 8-9 with columns 3, 5,
and 7 in Table S4 shows that the gain in power of using $l_{00}>0$ over
$l_{00}=0$ can be large.
Figure \ref{figpower} shows the average power and the power for at
least one of our proposal as a function of $c_2\in\{0.05, 0.1,
\ldots,0.95\}$ and $l_{00}\in\{0, 0.5, 0.8, 0.9\}$ for $q=0.05.$ As
expected, both measures of power increase as $l_{00}$ increases. For fixed $l_{00}$ and $(\pi_1, \pi_2)$, the highest average
power among all the choices of $c_2$ is close to the average power
when $c_2=0.5$ (Figure \ref{figpower}, left column), also
shown in Table S4. The power curve
for at least one as a function of $c_2$ is flat around $c_2=0.5$ (Figure \ref{figpower}, right column), suggesting as well that $c_2=0.5$ is an appropriate choice.

\begin{figure}[!tpb]\
\centering \subfloat[$\pi_1=0.1$,
$\pi_2=0.2$]{\label{fig20025:a}\includegraphics[width=0.33\textwidth,
height=5.5cm]{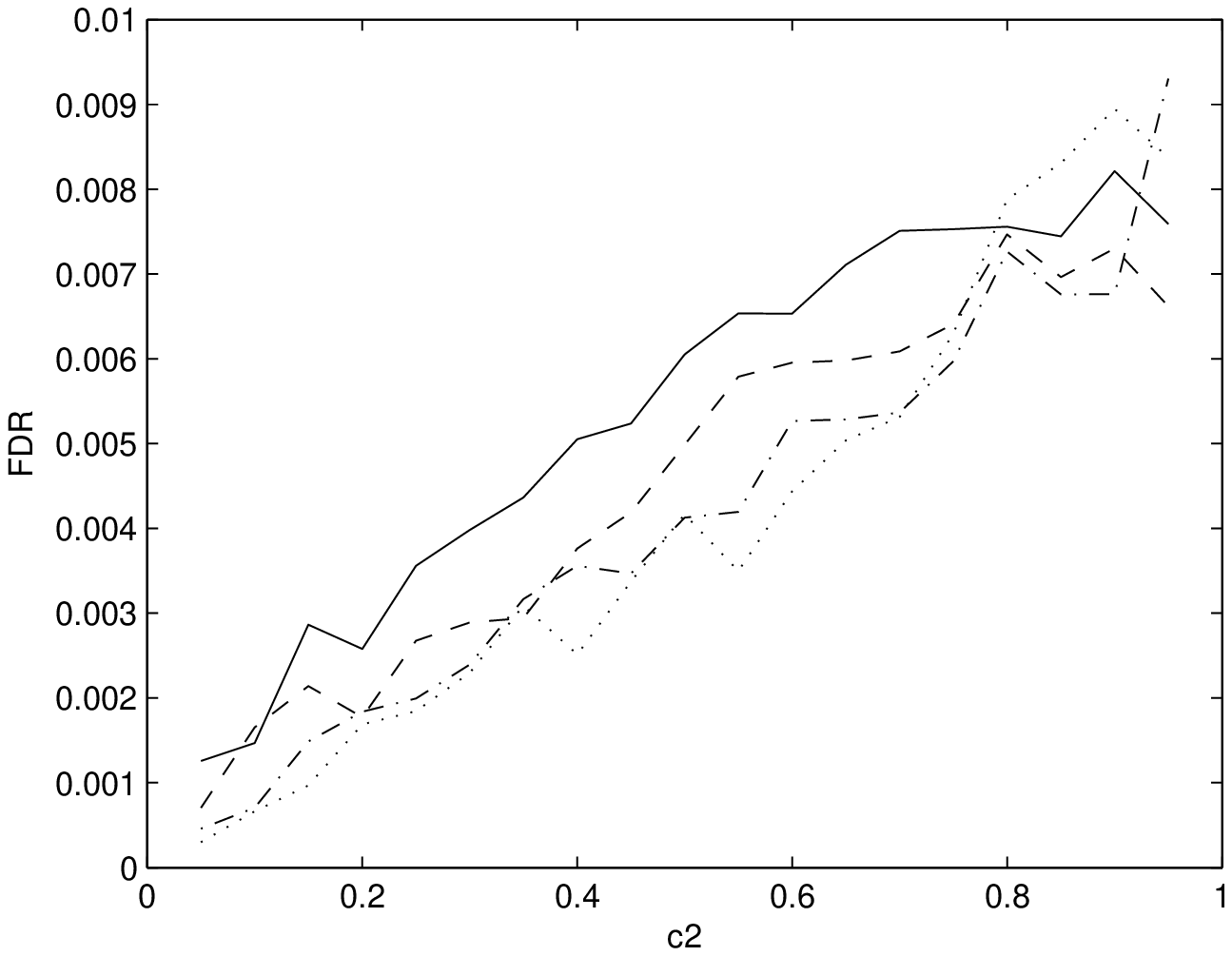}} \subfloat[$\pi_1=0.1$,
$\pi_2=0.5$]{\label{fig20025:b}\includegraphics[width=0.33\textwidth,
height=5.5cm]{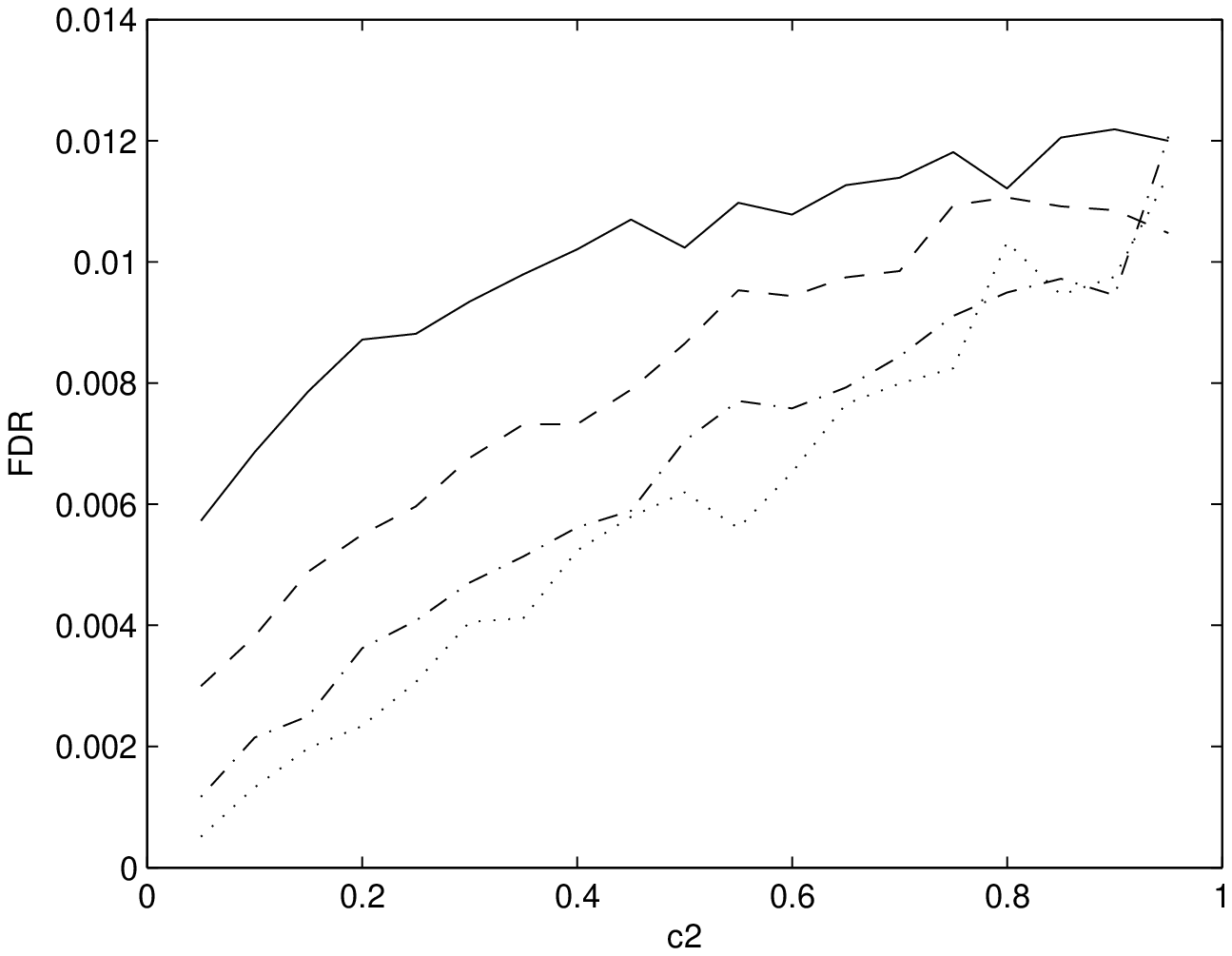}} \subfloat[$\pi_1=0.1,
\pi_2=0.8$]{\label{fig20025:c}\includegraphics[width=0.33\textwidth,
height=5.5cm]{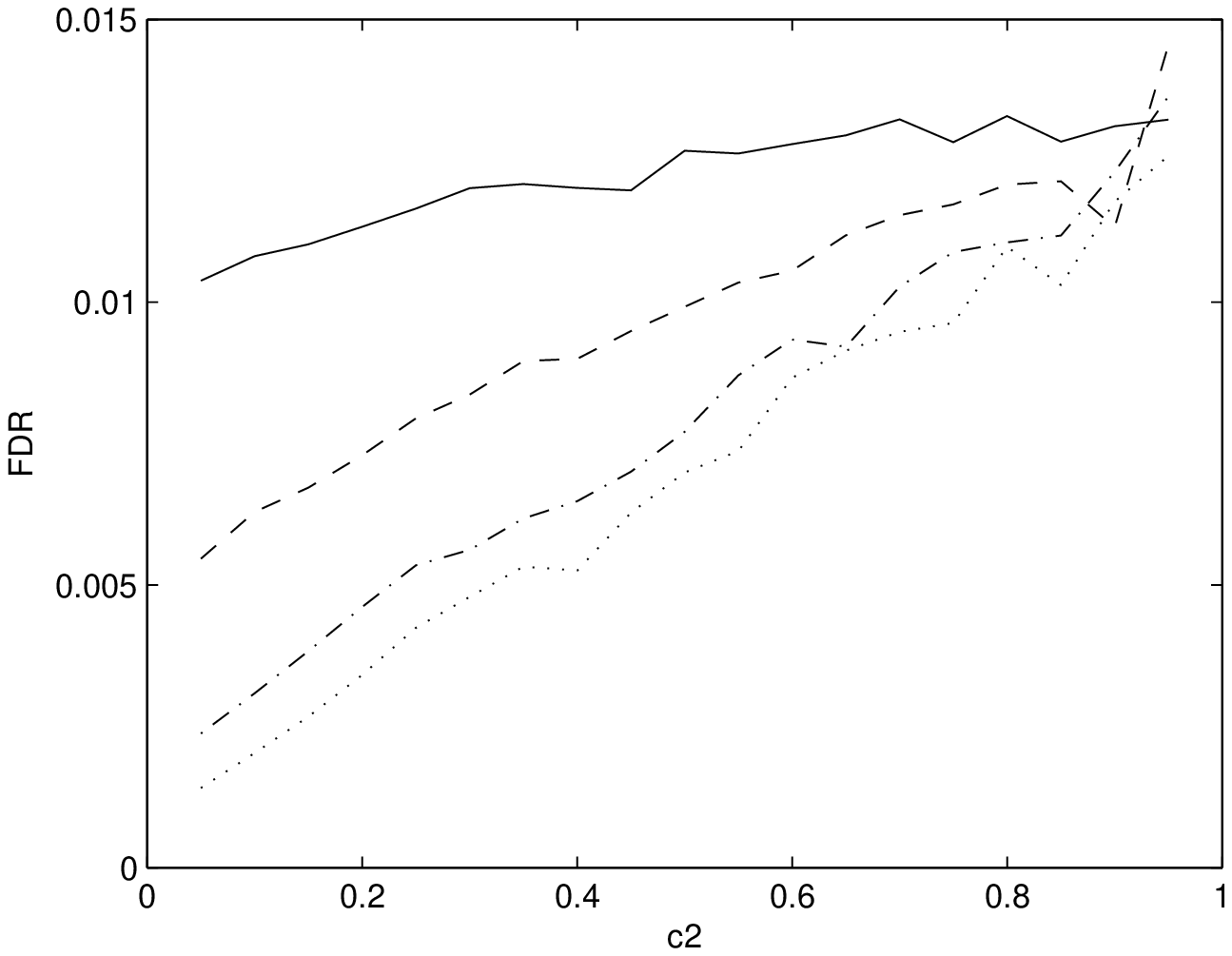}} \caption{The estimated FDR of our
proposal with parameters $(l_{00}, c_2,0.05)$ as a function of $c_2$
in a simulation where $f_{00}=0.9, f_{01}=f_{10}=0.025,
f_{11}=0.05,$ the number of hypotheses examined in the primary study
is 1000, and the signal to noise ratios for the primary study and
the follow-up study, $\mu_1/\sigma_1$ and $\mu_2/\sigma_2$,
respectively, are taken according to the requirement that the power
of the Bonferroni procedure at level 0.05 in the primary study is
$\pi_1,$ and in the follow-up study is $\pi_2$ for $(\pi_1,
\pi_2)=(0.1, 0.2)$ (left panel), $(\pi_1, \pi_2)=(0.1, 0.5)$ (middle
panel), $(\pi_1, \pi_2)=(0.1, 0.8)$ (right panel); $l_{00}=0.9$
(solid), $l_{00}=0.8$ (dashed), $l_{00}=0.5$ (dash-dotted), and
$l_{00}=0$ (dotted). The standard errors were of the order of
$10^{-3}$ or $10^{-4}$ for all the sets of
parameters.}\label{figfdr}
\end{figure}

Figure \ref{figfdr} shows the FDR of our proposal as a
function of $c_2\in\{0.05, 0.1, \ldots,0.95\}$, for $l_{00}\in\{0, 0.5, 0.8, 0.9\}$ and $q=0.05.$ %For most
%of the values of $c_2$ and $(\pi_1, \pi_2),$ FDR is an increasing
%function of $l_{00}.$ IT can be argued that FDR Since the
%differences between the functions are of the order of standard
%errors, it ca
It can be seen that the FDR is far below 0.05 for all the sets of
parameters considered. This follows from the fact that our data
generation may result in FDR much lower than the upper bound given
in (\ref{eq-FDRbound}).  In order to see this, note that it follows
from the proof of Theorem 1 that the FDR of our proposal achieves
the upper bound in (\ref{eq-FDRbound}) when the $p$-values under the
alternative are practically zero. In our simulation setting, this
condition would hold if $\mu_i,$ for $i\in\{1,2\}$ were always
extremely large when compared to $N(0,1)$ random variables, e.g.
$\mu_i\geq 4.$ Obviously this does not hold for our data generation
process. Therefore we could get higher FDR values for another data
generation process, however we still would not expect to achieve
0.05 because of using conservative upper bounds for $f_{01}$ and
$E(|I_{10}\cap \mathcal{R}_1|/(\max|\mathcal{R}_1|, 1))$ in
expression (\ref{eq-FDRbound}).

\begin{table}
\scriptsize
\caption{The estimated average power of our proposal with parameters
$(l_{00}, c_2,0.05),$ where $c_2$ is the optimal choice among the
values in $\{0.05, 0.1,\ldots, 0.95\}$ for $l_{00}=0.5$ (column 2),
$l_{00}=0.8$ (column 4), $l_{00}=0.9$ (column 6), and $l_{00}=0$
(column 8), the optimal value of $c_2$ is given in the row below;
$c_2=0.5,$ for $l_{00}\in\{0.5, 0.8, 0.9, 0\}$ (columns 3, 5, 7, 9)
 in a
configuration $f_{00}=0.9, f_{01}=f_{10}=0.025, f_{11}=0.05.$ The
number of hypotheses examined in the primary study is 1000. The
signal to noise ratios for the primary study and the follow-up
study, $\mu_1/\sigma_1$ and $\mu_2/\sigma_2$, respectively, are
taken according to the requirement that the power of Bonferroni
procedure at level 0.05 in the primary study is $\pi_1,$ and in the
follow-up study is $\pi_2$ (given in the first column). The standard
errors were of the order of $10^{-3}$ or $10^{-4}$ for all the
estimates.}
 \label{simtab1}
\begin{tabular}{|c|c|c|c|c|c|c|c|c|}
  \hline
   $(\pi_1, \pi_2)$& Optimal for  &  $l_{00}=0.5$   & Optimal for  & $l_{00}=0.8$ & Optimal for &$l_{00}=0.9$ & Optimal for  & $l_{00}=0$\\
                   &$l_{00}=0.5$ & $c_2=0.5$& $ l_{00}=0.8$&
               $c_2=0.5$ &  $l_{00}=0.9$ & $c_2=0.5$ & $l_{00}=0$ & $c_2=0.5$\\\hline%\\\hline
              &   & & & & & & & \\
  (0.1, 0.8) & 0.2980 & 0.2515 & 0.4486 & 0.3858 & 0.5681 & 0.4921 & 0.2009  & 0.1686 \\
             & $c_2=0.2$       &       &  $c_2=0.2$      &         &
            $c_2=0.15$ & &$c_2=0.2$&
              \\
  $(0.1, 0.5)$ & 0.1749 &  0.1666 & 0.2881& 0.2750 & 0.3837&
  0.3669 & 0.1105 & 0.1044 \\
     &$c_2=0.35$ & & $c_2=0.35$ & & $c_2=0.35$ & &$c_2=0.35$  &\\%\\\hline
   $(0.1, 0.2)$ & 0.0425 &  0.0425 & 0.0786& 0.0781 & 0.1152&
  0.1152 & 0.0261& 0.0258\\
     &$c_2=0.5$ & & $c_2=0.55$ & & $c_2=0.5$ & &$c_2=0.55$ &\\\hline%\\\hline
\end{tabular}
\end{table}


\section{GWAS simulation example}
The goal of the simulation was threefold. First, to verify that the
FDR is controlled below the nominal level for realistic simulations
with GWAS type dependency, even if hypotheses with primary study
$p$-values above $c_1(q)q/m$ are followed-up. Second, to compare the
performance of our suggested proposal with the BH procedure on
maximum $p$-values. Third, to examine the effect of $l_{00}$ on the
power of the two procedures.

The information on $l_{00}$ is incorporated into the  BH procedure
on maximum $p$-values, to make the comparison fair, by performing the BH procedure at level
$q/(1-l_{00})$. It is straightforward to show that the FDR is
controlled at level at most $q$ for the BH procedure on the maximum
$p$-values at level $q/(1-l_{00})$, when the $p$-values within each
study are independent.

We simulated two GWAS from the simulator HAPGEN2
\cite{Su11}. The two studies were generated from two samples of the
HapMap project \cite{HapMap03}, a sample of 165 Utah residents with
Northern and Western European ancestry (CEU), and a sample of 109
Chinese in Metropolitan Denver, Colorado (CHD). In the
 CEU and CHD populations, respectively, 34 and 38 SNPs were set as disease SNPs with an increased multiplicative
relative risk of 1.2, and 18 of the disease SNPs were common to both
populations. Each study contained 4500 cases and 4500 referents. The
linkage disequilibrium (LD) across SNPs, as measured for the samples
in the HapMap project, was retained. Due to LD, the number of SNPs
associated with the phenotype in each study was larger than the
number of disease SNPs. See \cite{Bogomolov12} for the details of this simulation.

The CHD study was the primary study, and the CEU study was the
follow-up study. SNPs were selected for follow-up only if they
were discovered by the BH procedure at level $c_1(0.05)\times 0.05$. Table
\ref{tab-GWASsimulation} presents the average number of replicated
findings, as well as the average false discovery proportion (FDP),
for our proposal with $c_2=0.5$ and $q=0.05$, and the BH
procedure on maximum $p$-values at level $0.05/(1-l_{00}),$ for
different values of $l_{00}$. From columns 4 and 7 it is clear that
the FDR is controlled and that our proposal is actually conservative,
for all values of $l_{00}$. From a comparison of columns 2 and 5 it
is clear that our proposal is more powerful than the BH procedure on
maximum $p$-values. Finally, from comparisons of the rows it is
clear that the power increases as $l_{00}$ increases.

\begin{table}[ht]
\scriptsize \centering \caption{For 4500 cases and 4500 referents in
both studies, the average number of associated and disease SNPs
discovered (SE), and the average FDP (SE), for different values of
$l_{00}$. The actual value of $f_{00}$ was above 0.999. Results  are
given for our proposal with $c_2=0.05$ and $q=0.05$ in
columns 2-4, and for the BH procedure on maximum $p$-values at level
$0.05/(1-l_{00})$ in columns 5-7. SNPs were selected for follow-up only if they
were discovered by the BH procedure at level $c_1(0.05)\times 0.05$.}\label{tab-GWASsimulation}
\begin{tabular}{|c|ccc|ccc|}
  \hline
  & \multicolumn{3}{|c|}{FDR $r$-values$\leq 0.05$} & \multicolumn{3}{|c|}{BH Procedure on maximum $p$-values at level $0.05/(1-l_{00})$} \\
  &  \multicolumn{2}{|c}{\# Replicated findings} & FDP & \multicolumn{2}{|c}{\# Replicated findings} &
FDP\\
 $l_{00}$ &associated SNPs (SE) & disease SNPs (SE) & (SE)  &associated SNPs (SE) & disease SNPs (SE) & (SE) \\
  \hline
0 & 41.5 (5.3) & 8.3 (0.5) & 0.011 (0.011)& 29.2 (3.2) & 7.4 (0.4) & 0.000 (0.000)\\
  0.8 & 55.4 (5.3) & 9.3 (0.4) & 0.013 (0.013)& 39.0 (3.5) & 8.5 (0.5) & 0.000 (0.000) \\
  0.9 &  58.4 (4.9) & 9.6 (0.3) & 0.014 (0.014)& 42.8 (3.3) & 9.1 (0.4) & 0.000 (0.000)\\
  0.95 & 59.9 (4.5) & 9.7 (0.3) & 0.015 (0.014) & 46.1 (3.5) & 9.3 (0.3) & 0.000 (0.000)\\
  0.99 &  60.0 (4.6) & 9.7 (0.3) & 0.015 (0.014) & 50.8 (3.9) & 9.4 (0.3) & 0.000 (0.000)\\
   \hline
\end{tabular}
\end{table}

\section{Procedure for FWER control}
%\begin{theorem}\label{thm_FWER}The procedure for replicability analysis with parameters
%$(l_{00}, c_2, \alpha)$ where $0\leq l_{00}<1,$ $0<c_2<1,$ and
%$0<\alpha<1,$ which declares as replicated findings the set
%\[\{j: \max(mp_{1j}/c_1, |\mathcal{R}_1|p_{2j}/c_2)\leq \alpha, j\in
%\mathcal{R}_1 \},\] where $c_1=(1-c_2)/(1-l_{00}(1-c_2\alpha)),$
%controls the FWER at level at most $\alpha$ if $l_{00}\leq f_{00},$
%and the follow-up study $p$-values are independent of the primary
%study $p$-values.\end{theorem}
\begin{theorem}\label{thm_FWER}

A procedure that declares findings with Bonferroni $r$-values at
most $\alpha$ as replicated controls the FWER for replicability
analysis at level at most $\alpha$ if $l_{00}\leq f_{00}$ and the
follow-up study $p$-values are independent of the primary study
$p$-values.

\end{theorem}

\textbf{Proof of Theorem \ref{thm_FWER}.} %More generally, we shall
%prove that the above procedure controls the FWER at level which is
%smaller or equal to
%\begin{equation}c_1c_2\alpha^2f_{00}E\left(\frac1{\max(R_1, 1)}I(R_1>0)\right)+ f_{01}c_1\alpha+c_2\alpha
%E\left(\frac{|\mathcal{R}_1\cap I_{10}|}{\max(R_1,
%1)}\right).\label{upper_FWER}
%\end{equation}
%Since $E\left(\frac1{\max(R_1, 1)}I(R_1>0)\right)\leq 1,$ this upper
%bound is smaller or equal to the upper bound for the FDR of
%Procedure 1 with parameters $(l_{00}, c_2, \alpha)$, given in
%expression (\ref{eq-FDRbound}) with $q=\alpha.$ We showed in the
%proof of Theorem 1 that the expression in (\ref{eq-FDRbound}) is at
%most $q$ if $l_{00}\leq f_{00}.$ Therefore, if the upper bound in
%(\ref{upper_FWER}) holds and $l_{00}\leq f_{00},$ Theorem
%\ref{thm_FWER} follows.
It is easy to show that the procedure that declares findings with
Bonferroni $r$-values at most $\alpha$ as replicated is equivalent
to that of declaring as replicated all features with $f_j^{Bonf}(\alpha)\leq
\alpha$. The equivalence follows from the facts that $f_j^{Bonf}(x)$ is a continuous function of $x$
and $f_j^{Bonf}(x)/x$ is strictly monotone decreasing.
We shall prove
that the above procedure controls the FWER at level which is smaller
or equal to
\begin{equation}c_1c_2f_{00}\alpha^2+ f_{01}c_1\alpha+c_2\alpha
E\left(\frac{|\mathcal{R}_1\cap I_{10}|}{\max(|\mathcal{R}_1|,
1)}\right),\label{upper_FWER}
\end{equation}
where $c_1=(1-c_2)/\left(1-l_{00}(1-c_2\alpha)\right).$ Note that this upper bound is equal to the upper bound given in
expression (\ref{eq-FDRbound}) with $q=\alpha.$ We showed in the
proof of Theorem 1 that the expression in (\ref{eq-FDRbound}) is at
most $q$ if $l_{00}\leq f_{00}.$ Therefore, if the upper bound in
(\ref{upper_FWER}) holds and $l_{00}\leq f_{00},$ Theorem
\ref{thm_FWER} follows.

 We shall now prove that the expression in (\ref{upper_FWER}) is an
upper bound for the FWER for replicability analysis, which is
$\textmd{Pr}(R_{00}+R_{10}+R_{01}>0).$ Note that
\begin{align*}
\textmd{Pr}(R_{00}+R_{10}+R_{01}>0)\leq
E(R_{00}+R_{10}+R_{01})=\sum_{x\in \{00,01,10 \}}\sum_{j\in I_x}
E\left(R_j\right).
\end{align*}
For the procedure that declares as replicated all features with
$f_j^{Bonf}(\alpha)\leq \alpha,$ which is equivalent to the
procedure that declares findings with Bonferroni $r$-values at most
$\alpha$ as replicated (as discussed above),
\begin{align}\label{eq1fwer}
E\left(R_j\right)=\textmd{Pr}\left(j\in \mathcal{R}_1, P_{1j}\leq
\frac{c_1\alpha}{m}, P_{2j}\leq
\frac{c_2\alpha}{\max(|\mathcal{R}_1|, 1)}\right).
\end{align}
We shall give an upper bound for expression (\ref{eq1fwer})  for
$j\in I_{01}$, $j\in I_{10}$, and $j \in I_{00}$. For $j\in I_{01},$
\begin{eqnarray}
 \textmd{Pr}\left(j \in \mathcal{R}_1,
P_{1j}\leq \frac{c_1\alpha}{m}, P_{2j}\leq
\frac{c_2\alpha}{\max(|\mathcal{R}_1|, 1)}\right)  \leq
\textmd{Pr}\left( P_{1j}\leq \frac{c_1\alpha}{m}\right)
\leq\frac{c_1\alpha}{m}, \label{end01fwer}
\end{eqnarray}
where the last inequality follows from the fact that $P_{1j}$ is a
null-hypothesis $p$-value.

For $j\in I_{00}\cup I_{10}$ and an arbitrary fixed
$p_1=(p_{11},\ldots, p_{1m})$ such that $|\mathcal{R}_1(p_1)|>0,$
\begin{align}
E\left(R_j\,|\,P_1=p_1\right)&=  %\textmd{Pr}\left(j \in
%\mathcal{R}_1(p_1), p_{1j}\leq \frac{c_1\alpha}{m}, P_{2j}\leq
%\frac{c_2\alpha}{|\mathcal{R}_1(p_1)|}
%\,|\,P_1=p_1\right) \nonumber \\
%&& =
 I\left(p_{1j}\leq \frac{c_1\alpha}{m}, j\in \mathcal{R}_1(p_1)
\right)\textmd{Pr}\left( P_{2j}\leq
\frac{c_2\alpha}{|\mathcal{R}_1(p_1)|}\,
|\,P_1=p_1\right) \nonumber \\
&\leq \frac{c_2\alpha}{|\mathcal{R}_1(p_1)|}I\left(p_{1j}\leq
\frac{c_1\alpha}{m}, j\in \mathcal{R}_1(p_1)
\right),\label{nullpv1fwer}
\end{align}
where inequality (\ref{nullpv1fwer}) follows from the independence
of the $p$-values across the studies and the fact that $P_{2j}$ is a
null-hypothesis $p$-value. Using (\ref{nullpv1fwer}) we obtain the
upper bounds on expression (\ref{eq1fwer}) for $j\in I_{10}$ and for
$j\in I_{00}$. For $j\in I_{10},$ it follows that
\begin{eqnarray*}
&&E\left(R_j\,|\,P_1=p_1\right)
\leq\frac{c_2\alpha}{|\mathcal{R}_1(p_1)|} I\left(j \in
\mathcal{R}_1(p_1)\right),
\end{eqnarray*}
therefore
\begin{eqnarray}
&&E\left(R_j\right) \leq  c_2\alpha E\left(\frac{ I\left(j \in
\mathcal{R}_1\right)}{\max(|\mathcal{R}_1|,
1)}\right).\label{fin10fwer}
\end{eqnarray}
For $j\in I_{00}$, it follows that
\begin{eqnarray}E\left(R_j\right)\leq
c_2\alpha E\left[\frac{I\left(P_{1j}\leq \frac{c_1\alpha}{m}, j\in
\mathcal{R}_1(P_1) \right)}{\max(|\mathcal{R}_1(P_1)|, 1)}
\right]\leq c_2\alpha E\left[I\left(P_{1j}\leq
\frac{c_1\alpha}{m}\right)\right]\leq
c_2\alpha\frac{c_1\alpha}{m},\label{uncond1fwer}
\end{eqnarray}
where the last inequality follows from the fact that $P_{1j}$ is a
null-hypothesis $p$-value. From summing over the upper bounds (\ref{end01fwer}),
(\ref{fin10fwer}), and (\ref{uncond1fwer}) it thus follows that
\begin{align*}
FWER\leq E(R_{00}+R_{10}+R_{01})\leq c_1c_2f_{00}\alpha^2+
f_{01}c_1\alpha+c_2\alpha E\left(\frac{|\mathcal{R}_1\cap
I_{10}|}{\max(|\mathcal{R}_1|, 1)}\right).
\end{align*}

\

\end{document}